\documentclass[11pt]{article}

\usepackage[T1]{fontenc}
\usepackage[utf8]{inputenc}
\usepackage{newtxtext,newtxmath}          
\usepackage[margin=1in]{geometry}
\usepackage{amsmath}
\usepackage{graphicx}
\usepackage{booktabs}
\usepackage{array}
\usepackage[round]{natbib}                 
\usepackage{xurl}
\usepackage[colorlinks=true,linkcolor=blue,citecolor=blue,urlcolor=blue]{hyperref}
\usepackage{caption}
\usepackage{float}
\usepackage{flafter}
\usepackage[section]{placeins}
\AddToHook{cmd/subsection/before}{\FloatBarrier}
\usepackage{microtype}
\graphicspath{{figures/}}

\newcommand{\degr}{$^{\circ}$}

\begin{document}

\title{\textbf{Evaluation of Machine-Learning Weather Models in Reproducing Mesoscale Tornado-Favorable Convective Environments}}
\author{%
  Yiding Luo$^{1}$, Rui Yuan$^{1}$, Shetian Xia$^{1}$, Royce Lee$^{1}$,
  and Alexandra Anderson-Frey$^{1,*}$\\
  {\small $^{1}$Department of Atmospheric and Climate Science, University of Washington, Seattle, Washington, USA}\\
  {\footnotesize $^{*}$Corresponding author: Alexandra Anderson-Frey, akaf@uw.edu}%
}
\date{}

\maketitle

\begin{abstract}
Machine-learning (ML)-based weather prediction models have developed rapidly in recent years and are already being used in operational weather forecasting. These ML-based models' performance has exceeded that of traditional numerical weather prediction (NWP) methods in certain implementations, with substantially lower computational cost. However, most ML-based models were trained on global data at a horizontal grid spacing of 0.25\degr, which may be too coarse to capture tornado-favorable parameters. The ability of these ML-based models to predict the parameters relevant for tornado-favorable convection remains unclear. Here, we compare the performance of three ML-based models---Pangu-Weather, FengWu (versions 1 and 2), and FuXi Weather---in forecasting ten distinct convective events associated with strong tornadoes that occurred in Oklahoma, USA. The proposed methodology incorporates five lead times, ranging from 24 to 168 h, for each event and uses reanalysis data as both model input and ground truth. We evaluate the models by forecasting eight tornado-relevant parameters and by conducting spatial and value-consistency checks between model outputs and ground truth. Performance varies across parameters, models, and events. Pangu-Weather performs well for positive storm-relative helicity, while FuXi Weather performs well for several thermodynamic parameters, albeit with some overprediction of magnitude. Increasing lead time is associated with loss of extreme-value and spatial-structure fidelity. In the Global Forecast System (GFS) case study, ML-based models retain convective signals at longer lead times but poorly represent fine-scale structures such as cold pool-like features. These findings indicate that model-specific biases and spatial extremes should be assessed alongside domain-wide forecast scores.
\end{abstract}

\section{Introduction}

Severe convective weather events, particularly those capable of producing tornadoes, present a serious threat to life and property. For instance, the Moore tornado of 20 May 2013 (rated EF5 on the Enhanced Fujita (EF) scale) caused massive damage in Oklahoma, resulting in 24 fatalities and more than 200 injuries~\citep{burgess20May20132014}, with estimated property damage of approximately \$2 billion~\citep{StormEventsDatabasea}. Accurately forecasting the environments that support these hazardous phenomena, especially for tornadoes, is a major open challenge in mesoscale meteorology~\citep{stensrudProgressChallengesWarnonForecast2013,huaPerformancePanguWeatherDeep2025}.

Historically, operational forecasting has relied on Numerical Weather Prediction (NWP) models, such as the Global Forecast System (GFS), which simulate atmospheric dynamics by numerically solving governing dynamic and thermodynamic equations. While NWP models remain fundamental to forecasting pipelines, they are computationally intensive~\citep{stensrudProgressChallengesWarnonForecast2013} and sensitive to initial condition errors~\citep{lorenzPredictabilityFlowWhich1969}. These constraints often limit their ability to accurately resolve the highly localized, nonlinear environmental precursors necessary for mesoscale convective initiation~\citep{melhauserPracticalIntrinsicPredictability2012}, leading to decreased forecast skill at extended lead times, thereby limiting reliable tornado-environment forecasts beyond a few days.

In recent years, data-driven deep learning architectures have expanded the framework of atmospheric modeling. High-performance machine learning (ML) models trained on massive historical reanalysis datasets, such as the European Centre for Medium-Range Weather Forecasts (ECMWF) Reanalysis v5 (ERA5), have demonstrated an ability to generate global forecasts that match or exceed the accuracy of traditional NWP baselines at a fraction of the computational cost. The architectures of these ML-based models vary considerably, ranging from the early FourCastNet model, which is based on the Adaptive Fourier Neural Operator~\citep{pathakFourCastNetGlobalDatadriven2022}, to transformer-based models such as Pangu-Weather~\citep{biAccurateMediumrangeGlobal2023}. These models are continuously redefining forecast skill benchmarks. ECMWF, an established operational forecasting center, has also developed its own ML-based model, the Artificial Intelligence Forecasting System (AIFS), which is strongly competitive with leading NWP systems~\citep{langAIFSECMWFsDatadriven2024} and has already been applied in operational forecasting. Among these diverse architectures, transformer-based models represent the most architecturally varied subset, with distinct loss functions and training strategies whose impacts on convective-scale prediction remain largely unexplored. However, existing evaluations of ML-based models for convective environments remain limited.

\citet{feldmannLightningFastConvectiveOutlooks2024} took the first step toward evaluating the convective-environment prediction performance of ML-based models, demonstrating that GraphCast and Pangu-Weather can skillfully forecast Convective Available Potential Energy (CAPE) and deep-layer shear across multiple continents. However, their analysis was limited to three convective parameters (CAPE, deep-layer shear, and their product), which did not fully characterize the range of atmospheric conditions conducive to tornado initiation. Furthermore, while their evaluation incorporated the Fractions Skill Score (FSS) for spatial verification, it did not employ event-based categorical metrics or object-based diagnostic methods, limiting the assessment of localized extreme-value structures that are central to tornado prediction.~\citet{huaPerformancePanguWeatherDeep2025} conducted the first evaluation of Pangu-Weather's performance in predicting tornado-favorable environments across the contiguous United States, using a dataset of approximately 3{,}000 tornado events. Their analysis, however, focused solely on a single ML-based model at a fixed lead time of 24 hours, leaving cross-model comparisons and the evolution of model performance with extended lead times unexplored.

Despite their success in predicting large-scale synoptic features, the capacity of these global ML-based models to resolve localized, mesoscale environments favorable for severe convective outbreaks and tornadogenesis remains an open question. Severe weather forecasting relies on the accurate and timely derivation of diagnostic convective indices from modeled atmospheric profiles, including thermodynamic parameters such as CAPE, K-Index, and Lifted Condensation Level (LCL), kinematic parameters such as storm-relative helicity (SRH), and composite indices such as the Significant Tornado Parameter (STP)~\citep{thompsonCloseProximitySoundings2003}. A common thread emerging from these studies is that ML-based weather models systematically smooth forecast fields with increasing lead time, suppressing the extreme values that define severe convective environments.~\citet{bouallegueRiseDataDrivenWeather2024} characterized this oversmoothing as a fundamental limitation of current data-driven models, while~\citet{bonavitaLimitationsCurrentMachine2024} demonstrated that their effective spatial resolution is substantially lower than the nominal 0.25\degr{} grid spacing, rendering mesoscale features particularly vulnerable to attenuation. More recently,~\citet{zhangPhysicsbasedModelsOutperform2026} found that physics-based HRES forecasts outperformed several AI models for record-breaking heat, cold, and wind events. Their findings distinguish performance on record-breaking extremes from aggregate forecast skill and motivate evaluation of high-end convective environments. The forecast value of these convective parameters lies primarily in their spatial extremes: regions of elevated CAPE, helicity, and instability indices coincide with environments capable of supporting severe convection and tornadoes~\citep{brooksSpatialDistributionSevere2003}. Quantifying how this oversmoothing manifests across different ML architectures, parameters, and lead times, and how it interacts with model-specific internal biases, is therefore critical for assessing the operational readiness of ML-based models for severe weather forecasting.

Although oversmoothing is a well-documented characteristic of ML-based models, different architectures may show diverse oversmoothing behaviors~\citep{bonavitaLimitationsCurrentMachine2024}, whether in terms of spatial patterns or evolution with lead time~\citep{bouallegueRiseDataDrivenWeather2024}. Moreover, a single model may behave differently depending on the selected tornado-relevant parameters, leading to varying oversmoothing trajectories. Therefore, it is overly simplistic to assert that all ML-based models uniformly blur the forecast fields, and systematically characterizing the oversmoothing patterns across different models in convective environments remains a challenging task. To conduct a more comprehensive assessment of tornado event environments, diverse sources of atmospheric parameters, including thermodynamics, kinematics, cloud-relevant variables, and composite metrics such as STP~\citep{thompsonCloseProximitySoundings2003}, would be incorporated into model performance evaluation, but this incorporation would come at a relatively high computational cost. Furthermore, conventional continuous metrics such as root-mean-square error (RMSE), which are commonly employed in ML-based model evaluations, may obscure poor performance in extreme values and fail to explicitly characterize unique features like variance compression and lack of physical consistency~\citep{bonavitaLimitationsCurrentMachine2024}. As~\citet{roebberVisualizingMultipleMeasures2009} noted, ``it is essential to recall that there is no universal approach to verification but rather that the procedure selected needs to match the specific objectives of the study.'' Consequently, a well-designed and holistic performance evaluation system must be established, particularly for extreme weather events.

This study evaluates the performance of three transformer-based ML weather prediction models: Pangu-Weather, FengWu (versions 1 and 2), and FuXi Weather, alongside the operational GFS. We assess their predictive skill for ten historically significant, strong tornado event environments located within the highly active corridor of central Oklahoma, USA~\citep{brooksSpatialDistributionSevere2003}. We assess eight tornado-relevant parameters spanning thermodynamic diagnostics (CAPE, convective inhibition (CIN), K-Index), kinematic diagnostics (0--3 km positive and negative storm-relative helicity (SRH)), cloud-related diagnostics (lifted condensation pressure level (LCPL), Pressure Level of Free Convection (PLFC)), and the composite significant tornado parameter (STP). Using five distinct forecast lead times ranging from 24 to 168 hours, model outputs are verified against the ERA5 reanalysis, which serves as both the model input and the ground truth for all ML-based models. To evaluate the value and spatial consistency, as well as biases, of the tornado-relevant parameters, we employ the performance diagram~\citep{roebberVisualizingMultipleMeasures2009} and its associated metrics, including Probability of Detection (POD), Success Ratio (SR), Frequency Bias (FB), and Critical Success Index (CSI). For geometric and positional fidelity of high-value convective objects, we further introduce the Fractions Skill Score (FSS)~\citep{robertsScaleSelectiveVerificationRainfall2008} and the total-interest score from the Method for Object-Based Diagnostic Evaluation (MODE)~\citep{davisMethodObjectBasedDiagnostic2009}. This framework complements continuous metrics such as RMSE by directly evaluating the spatial correspondence and structure of extreme-value regions, rather than averaging over the entire forecast domain. Through this rigorous comparative framework, we aim to clarify the strengths, biases, and operational potential of deep learning-based weather models in high-impact, mesoscale forecasting contexts. Specifically, we show that the oversmoothing behavior is architecture-dependent, that the evaluated models exhibit distinct model-specific internal biases in their convective environment predictions, and that the ML-based models differ fundamentally from the NWP baseline GFS in how forecast skill degrades with lead time.

\section{Methodology}
\subsection{Data preparation}

\begin{figure}[htbp]
  \centering
  \includegraphics[width=0.95\textwidth]{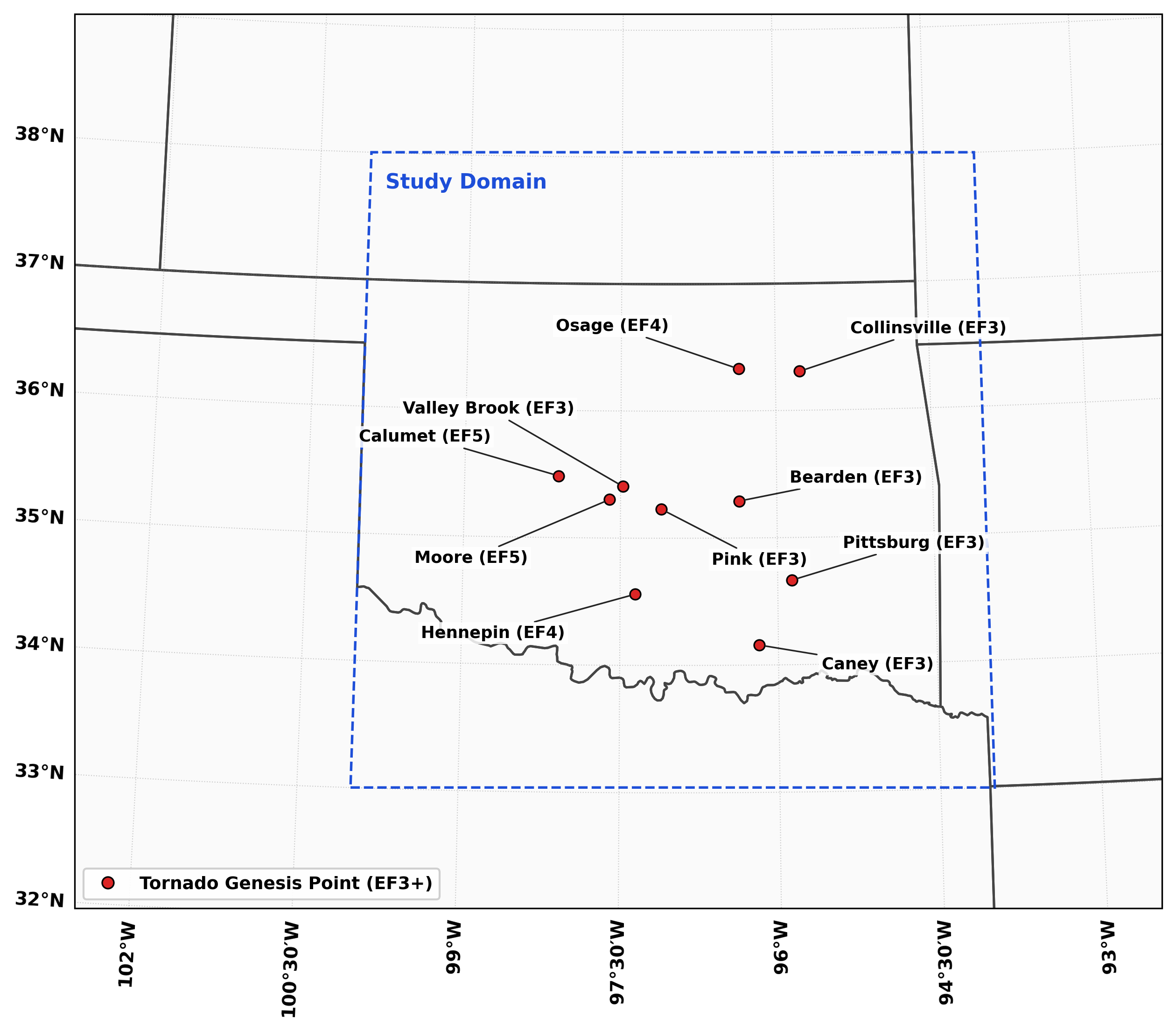}
  \caption{Visualization of the study region (central Oklahoma, USA). The dashed box outlines the study domain (100\degr W--94\degr W, 33\degr N--38\degr N), and red dots mark the tornado genesis points of the ten EF3+ events examined in this study, labeled with the NOAA-reported location and EF rating given in Table~\ref{tab:events}.}
  \label{fig:study_area}
\end{figure}

The ten selected severe convective environments correspond to timestamps that include EF3+ NOAA-confirmed tornadoes~\citep{StormEventsDatabase}. Since the ECMWF Reanalysis v5 (ERA5) data~\citep{hersbachERA5GlobalReanalysis2020} are only available at the top of each hour, the tornado's recorded start time is rounded to the nearest hour, which is then called the event time. For the ten events considered, the time differences between the event time and the reported tornado start time do not exceed 20 minutes. The tornado event environments considered span the period from 2013 to 2025, and the locations of the tornadoes are restricted to the central Oklahoma, USA region (Fig.~\ref{fig:study_area}). The ten selected event environments are listed in Table~\ref{tab:events}.

\begin{table}[htbp]
  \centering
  \caption{Examined tornado event environments from the NOAA Storm Events Database~\citep{StormEventsDatabase}. Location names follow the database reports and are distinct from the county/zone field. EF denotes the Enhanced Fujita scale; all listed locations are in Oklahoma, USA.}
  \label{tab:events}
  \small
  \begin{tabular}{@{}llll@{}}
    \toprule
    Start Time (UTC) & Coordinates (Lat/Lon) & NOAA-reported location & EF Scale \\
    \midrule
    May 20, 2013, 20:04 & 35.303 / $-97.605$ & Moore          & EF5 \\
    May 31, 2013, 23:03 & 35.485 / $-98.096$ & Calumet        & EF5 \\
    May 07, 2015, 01:41 & 35.406 / $-97.477$ & Valley Brook   & EF3 \\
    May 25, 2015, 20:10 & 34.157 / $-96.179$ & Caney          & EF3 \\
    May 09, 2016, 21:06 & 34.559 / $-97.357$  & Hennepin       & EF4 \\
    April 20, 2023, 02:04 & 35.229 / $-97.111$ & Pink           & EF3 \\
    April 28, 2024, 04:04 & 35.290 / $-96.363$  & Bearden        & EF3 \\
    May 07, 2024, 02:12 & 36.330 / $-96.356$ & Osage          & EF4 \\
    May 26, 2024, 04:19 & 36.307 / $-95.765$ & Collinsville   & EF3 \\
    May 19, 2025, 22:44 & 34.665 / $-95.863$  & Pittsburg      & EF3 \\
    \bottomrule
  \end{tabular}
\end{table}

The input data for all included ML-based models, as well as the reference fields for the tornadic convective environment, are derived from ERA5 reanalysis. While these reference fields do not represent the true atmospheric state, they are used as the ground truth for model consistency checks to ensure fairness. The included ML-based models are all based on transformer architectures, but each incorporates distinct architectural modifications. They are Pangu-Weather~\citep{biAccurateMediumrangeGlobal2023}, FengWu~\citep{chenOperationalMediumrangeDeterministic2025}, and FuXi Weather~\citep{chenFuXiCascadeMachine2023}. Note that the second version of FengWu is designed primarily for operational forecasting with transfer learning trained from FengWu V1~\citep{chenOperationalMediumrangeDeterministic2025}, yet its performance under ERA5 reanalysis conditions remains of interest for comparative analysis.

The model input and inference processes have strictly adhered to the guidelines provided by the model developer or the affiliated organization for each model. All the models require global surface variables (2m temperature, mean sea level pressure, 10m u-component of wind, and 10m v-component of wind) and variables at 13 global pressure levels (temperature, geopotential height, humidity, u-component of wind, v-component of wind) as the input data. The pressure levels are (1000, 925, 850, 700, 600, 500, 400, 300, 250, 200, 150, 100, 50 hPa). However, Pangu-Weather and FengWu require specific humidity, while FuXi Weather requires relative humidity and an additional surface variable, total precipitation. Additionally, FengWu and FuXi Weather require two consecutive inputs with a 6-hour time difference.

\begin{figure}[htbp]
  \centering
  \includegraphics[width=0.88\textwidth]{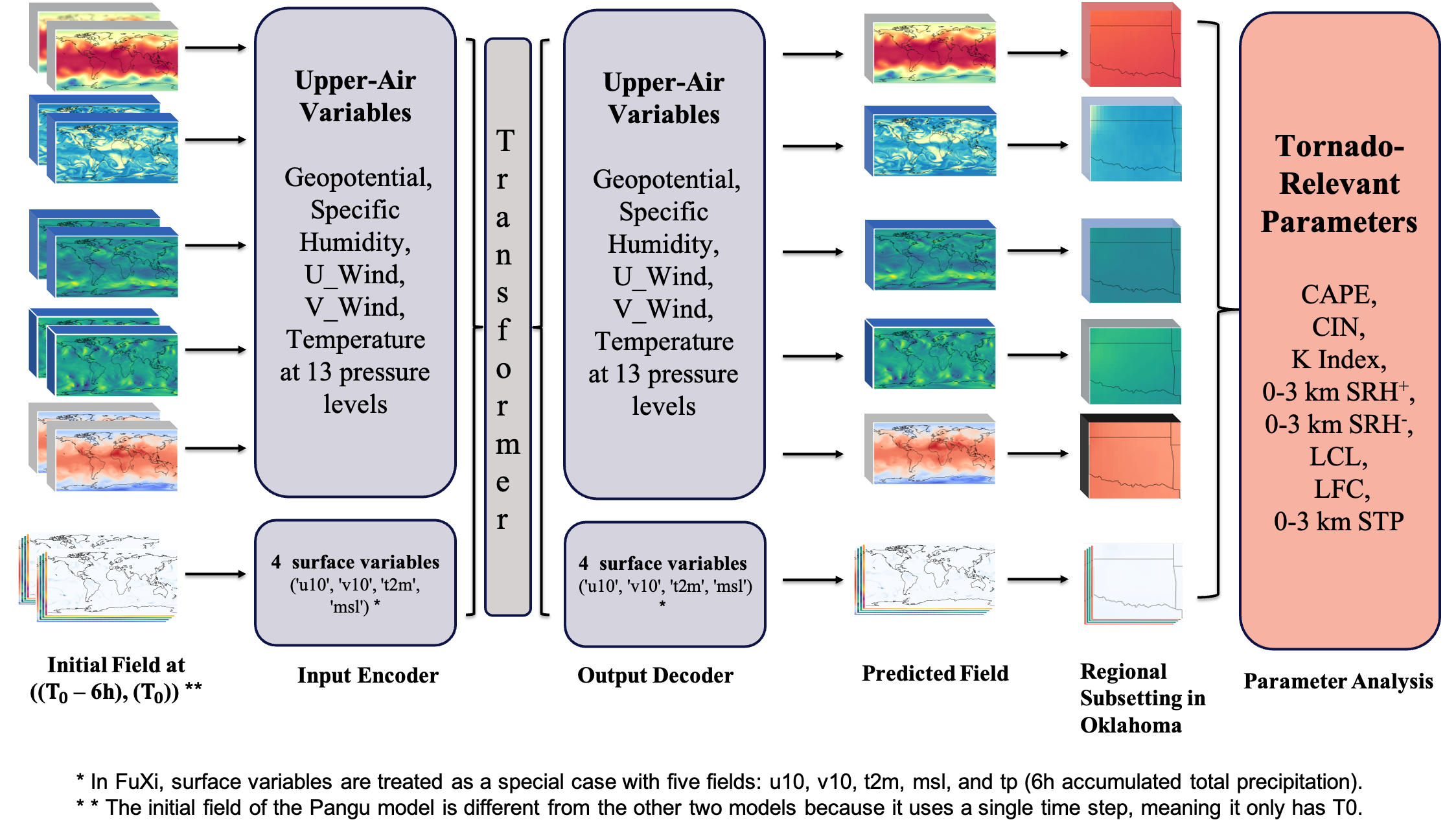}
  \caption{Flowchart of data ingestion, inference, and parameter calculation for the ML-based models. Here, $T_0$ is the forecast initialization time; u10 and v10 are the 10 m zonal and meridional wind components; t2m is 2 m temperature; msl is mean sea level pressure; and tp is total precipitation. U and V denote zonal and meridional wind. Parameter abbreviations are defined in Section~\ref{sec:parameter_calculations}; SRH$^+$ and SRH$^-$ denote the positive and negative 0--3 km SRH values.}
  \label{fig:dataflow}
\end{figure}

The lead times for each convective environment and for the model are 24, 48, 72, 120, and 168 hours. Accordingly, the input data consist of ERA5 reanalysis fields at these lead times before the event time. Since FengWu and FuXi Weather require two consecutive input timestamps, each model is initialized with the field six hours before the lead time followed by the lead-time field itself, in chronological order, which matches the models' requirements. The data ingestion and inference processes of the ML-based models are shown in Fig.~\ref{fig:dataflow}.

\begin{figure}[htbp]
  \centering
  \includegraphics[width=0.8\textwidth]{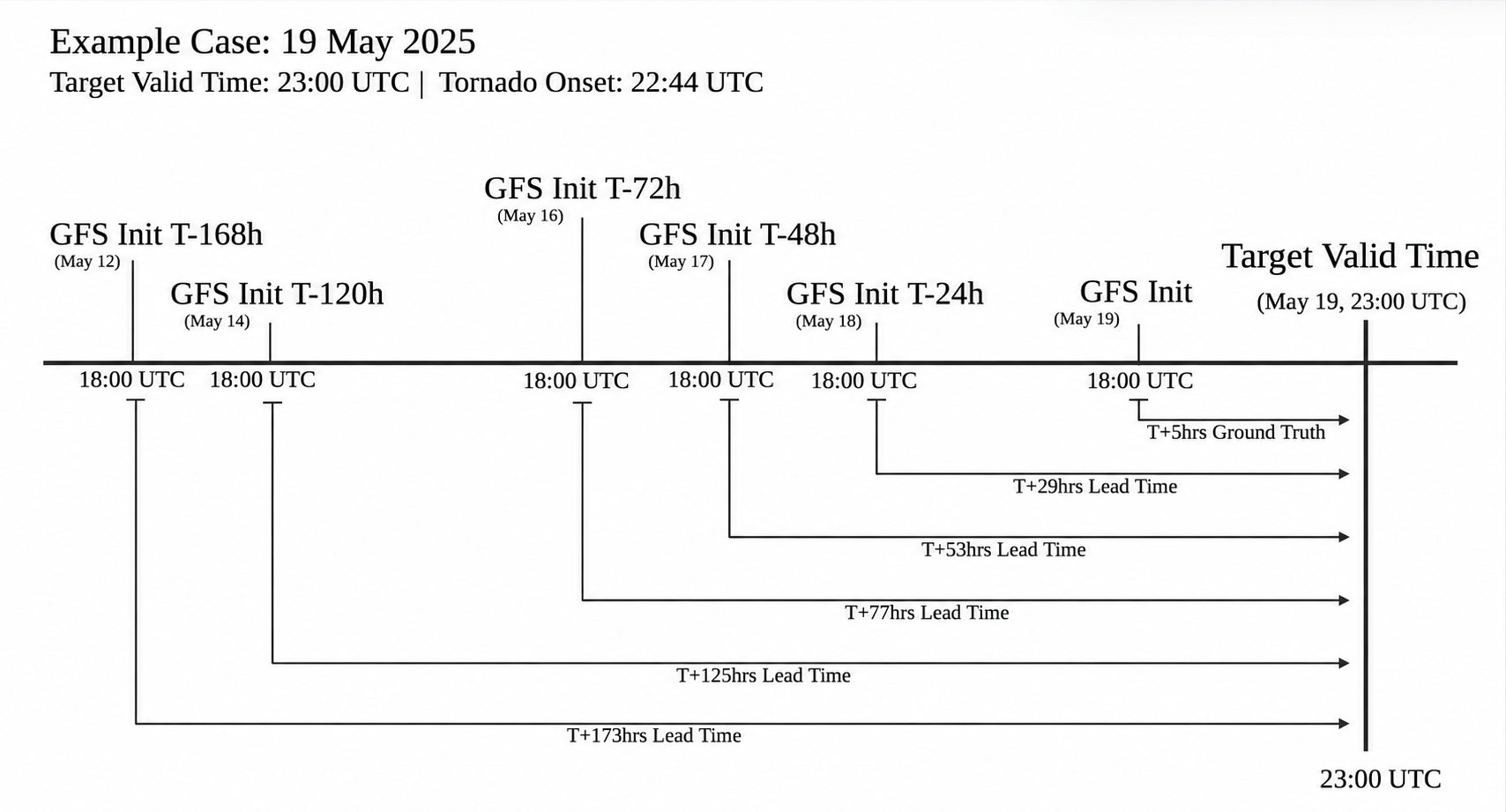}
  \caption{GFS initialization cycles and forecast-hour alignment for the 19 May 2025 tornado case. The tornado began at 22:44 UTC; the target valid time for environmental evaluation is 23:00 UTC. Here, $T$ denotes 18:00 UTC on 19 May, the reference GFS initialization cycle. A cycle initialized $\tau$ hours before $T$ requires a forecast hour of $\tau+5$ to reach the target valid time. Numbers on the arrows denote forecast durations from their respective initialization cycles. The reference field is the $+5$ h forecast from the reference cycle. The timeline is schematic and not to scale.}
  \label{fig:gfs_timeline}
\end{figure}

Furthermore, the Global Forecast System (GFS) is included as an NWP baseline for comparison~\citep{commerceNCEPGFS0252015}. Because GFS is initialized only four times per day (00, 06, 12, and 18 UTC), the initialization time closest to each target convective environment time is selected. When the target time does not coincide with a GFS initialization cycle, the forecast valid at the target time is obtained by adding the time difference between the selected initialization time and the target time to the lead time. This procedure is applied to all lead times to ensure temporal alignment between the forecasts and the target environment time. The ground truth in this case will be the forecast of GFS from the selected initialization time to the target convective environment time. The data selection process for GFS is shown in Fig.~\ref{fig:gfs_timeline}. Since GFS output is not available at hourly resolution before June 2019, the convective environment at the exact event hour cannot be captured for earlier events. Thus, a tornado event after June 2019 is selected to compare GFS with the ML-based models. The selected case study is an out-of-sample period beyond the training time range for all ML-based models.

\subsection{Tornado-relevant parameter calculations}
\label{sec:parameter_calculations}

To examine the convective environment, certain tornado-relevant parameters have been calculated. They are surface-based convective available potential energy (CAPE), surface-based convective inhibition (CIN), atmospheric K-Index, adjusted significant tornado parameter (STP), 0--3 km positive and negative storm-relative helicity (SRH), the pressure level associated with the lifted condensation level (here defined as the lifted condensation pressure level, LCPL), and the pressure level of free convection (here defined as PLFC). Since all models include only 13 pressure levels, the 0--1 km SRH may not be trustworthy, even though it is more relevant to tornado processes~\citep{huaPerformancePanguWeatherDeep2025}. The STP here will also use 0--3 km SRH and adopt surface-based (SB) parcel calculations for CAPE and LCL. We follow~\citet{thompsonCloseProximitySoundings2003} to adjust the SRH calculation in the STP formula from 0--1 km SRH to 0--3 km SRH with the denominator 100~m$^2$~s$^{-2}$ (Equation~\ref{eq:stp}):
\begin{equation}
  \label{eq:stp}
  \mathrm{STP} = \frac{\mathrm{SBCAPE}}{1500~\mathrm{J\,kg^{-1}}}
  \times \frac{2000 - \mathrm{SBLCL_{height}}}{1000~\mathrm{m}}
  \times \frac{\mathrm{SRH_{0-3\,km}}}{100~\mathrm{m^2\,s^{-2}}}
  \times \frac{\mathrm{Shear_{6\,km}}}{20~\mathrm{m\,s^{-1}}} .
\end{equation}
The SB approach has been validated for tornado-environment evaluation (e.g., the Storm Prediction Center (SPC) fixed-layer STP; the MetPy \texttt{significant\_tornado} implementation following~\citet{thompsonUpdateSupercellComposite2004}) and provides a consistent comparison across models since the same calculation method is applied to both forecasts and reanalysis. We acknowledge that mixed-layer parameters are physically preferred and may be able to capture elevated convection, but the operational use of SB parameters by SPC (the ``fixed-layer'' STP variant) supports the validity of this approach for supercell tornado environments.

All tornado-relevant parameter calculations are performed in the MetPy Python package~\citep{mayMetPyPythonPackage2025}. Since all model outputs are global, the parameter calculations are restricted to the Oklahoma, USA region of interest.

Since regional terrain can affect the calculation of certain parameters, such as surface-based CAPE and CIN, Shuttle Radar Topography Mission (SRTM) data are incorporated into the analysis~\citep{earthsciencedatasystemsNASAShuttleRadar2025}. The isothermal barometric formula is employed to approximately match the altitude of each grid point to its corresponding pressure (Equation~\ref{eq:pcalc}):
\begin{equation}
  \label{eq:pcalc}
  P_{\mathrm{calc}} = \mathrm{MSLP} \times e^{-\frac{z}{H}} ,
\end{equation}
where MSLP is the mean sea level pressure and $H$ is the scale height. This value is subsequently mapped to the nearest of the model's 13 discrete pressure levels, which is then defined as the surface-representative level for parameter calculation (Equation~\ref{eq:ptarget}):
\begin{equation}
  \label{eq:ptarget}
  S = \{1000, 925, 850, \dots\}, \qquad
  P_{\mathrm{target}} = \arg\min_{p \in S} \left| p - P_{\mathrm{calc}} \right| .
\end{equation}
In addition, SRTM data are used in SRH calculations to determine the height above ground level (AGL).

\subsection{Model performance evaluation techniques}

The evaluation of the models' performance is based on eight tornado-relevant parameters derived from the model output variables. To quantify the performance, metrics such as Probability of Detection (POD), Critical Success Index (CSI), Success Ratio (SR), and Frequency Bias (FB) are introduced~\citep{roebberVisualizingMultipleMeasures2009}. To focus on crucial convective regions, binarization is employed in the analysis. For all parameters except CIN and negative SRH, the threshold for binarization is set at the 75th percentile (P75) of the reference field, thereby isolating the upper tail of the distribution where higher values generally indicate enhanced tornado potential. For CIN and negative SRH, however, tornadic-favorable conditions are characterized by small negative values. Adopting the 75th percentile in these cases would produce a nearly uniform field of detections after binarization, rendering model performance indistinguishable. Therefore, to instead assess the models' fidelity in reproducing extreme magnitudes, the threshold for CIN and negative SRH is taken as the 25th percentile, focusing on the most negative values in the domain. In addition, LCPL and PLFC are expressed in hectopascals (hPa); therefore, larger values correspond to lower geopotential heights, which favor tornadic environments. Performance diagrams are used to visualize and examine all metrics listed above. The definitions of POD, CSI, SR, FB, and the other verification metrics used in this study are given in Appendix~A.

The spatial evaluation metrics are the Fractions Skill Score (FSS) and the Method for Object-based Diagnostic Evaluation (MODE). Both use reference-field percentile thresholds: P75 for most parameters and P25 for CIN and negative SRH. The neighborhood scale in FSS is a $3\times3$ grid. Since FSS is highly sensitive to its neighborhood size, we followed~\citet{antonioHowDeriveSkill2025} to evaluate forecast skill by comparing the FSS against its random forecast baseline, FSS$_{\mathrm{random}}(n)$, at multiple neighborhood scales ($n = 0, 3, 5, 7, 11$ grids). Among all eight tornado-relevant parameters and lead times, the gap between the forecast FSS and FSS$_{\mathrm{random}}$ is largest at $n = 3$ grids, indicating that the model shows the greatest skill relative to a random forecast at this scale. Conversely, as the neighborhood size increases, the FSS approaches the random baseline, reducing its differentiating skill despite the numerically higher raw FSS value. MODE is a spatial analysis method designed to identify the geometric shape and orientation of high-end value regions within a forecast field, and to quantify discrepancies in centroid position, area ratio, and orientation angle relative to a reference field. MODE maps differences in object attributes to interest values between 0 and 1 and combines them into a total-interest score, with larger scores indicating greater similarity. Interest values are calculated according to the Model Evaluation Tools (MET) documentation~\citep{20MODETool}. To avoid unreasonable forced object matches, the lowest threshold value of the total interest is 0.7. Any object pair with a total interest value below 0.7 is considered an unreasonable match, and its total interest value is therefore set to 0~\citep{davisMethodObjectBasedDiagnostic2009}.

All metrics are first calculated for each of the ten tornado event environments and then averaged across environments for each model. Cross-case error bars use a common dispersion-selection procedure: sample standard deviation (SD), trimmed SD, or a robust scale based on the median absolute deviation (MAD) or interquartile range (IQR), as detailed in Appendix~\ref{app:dispersion}. Each bar extends by one selected dispersion scale on either side of the arithmetic mean, subject to the plotting limits. These bars describe variability across environments, not standard errors or confidence intervals for the mean. Spatial SD within an individual field remains part of the field-variability metrics and is distinct from cross-case dispersion. The averaging process does not apply to the single-case analysis that includes the GFS comparison.

\section{Results}

\subsection{Metrics and performance diagram}
\label{sec:results_perf}

The following figures show CAPE, K Index, positive 0--3 km SRH, LCPL, PLFC, and STP. Complete eight-parameter results, including CIN and negative SRH, are provided in the \href{https://github.com/Mixtures-0218/Examine-ML-based-Models-Tornadic-Environment-Prediction-Performance/tree/f52f9474343860cebb6028e6d2878356a9c7fc3e/figures/supplementary}{repository supplementary figures (S1--S7)}.

\begin{figure}[p]
  \centering
  \includegraphics[width=\textwidth,height=0.87\textheight,keepaspectratio]{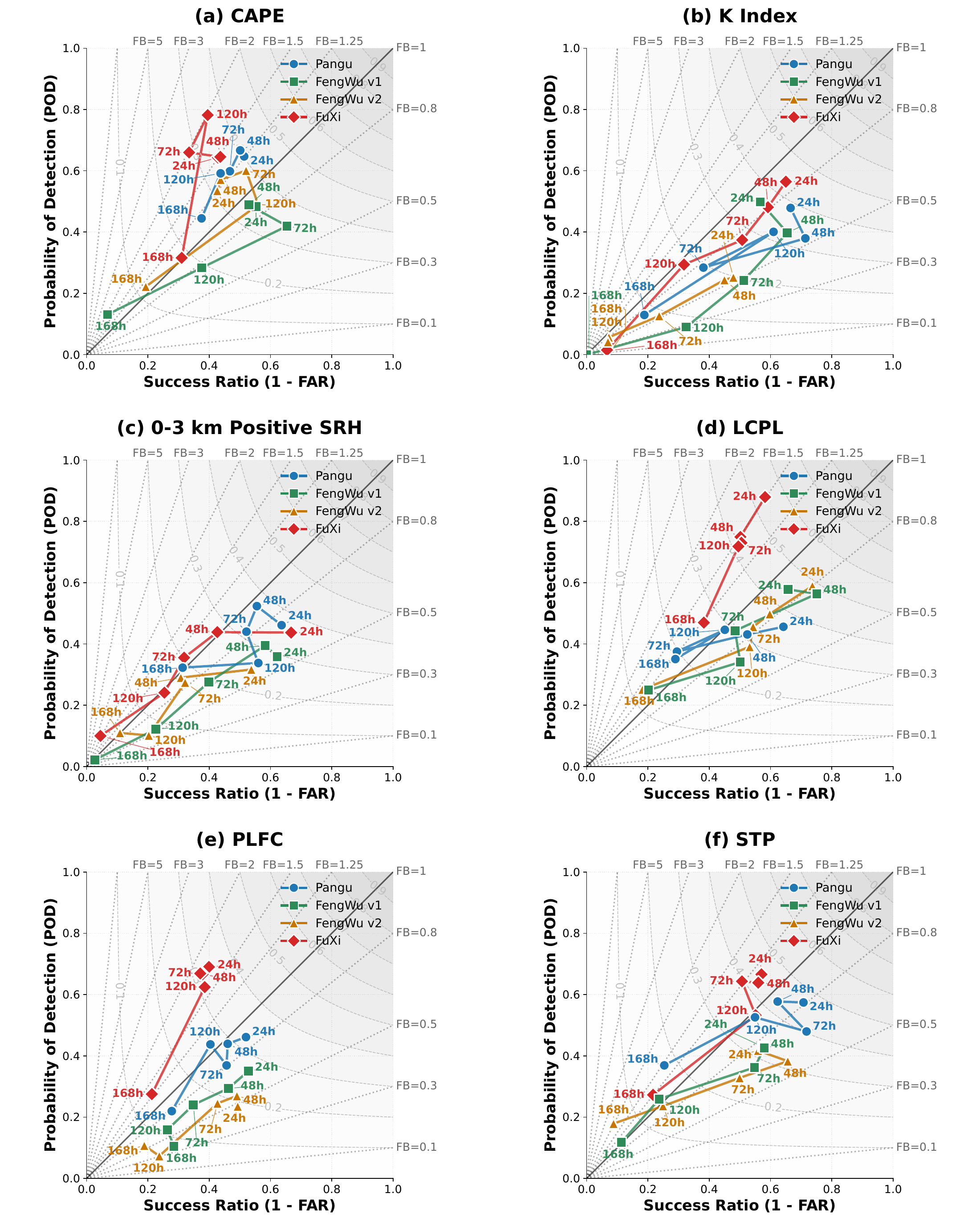}
  \caption{Mean performance diagrams for six tornado-relevant parameters across ten tornado events: (a) CAPE, (b) K Index, (c) positive 0--3 km SRH, (d) LCPL, (e) PLFC, and (f) STP. Colors and marker shapes identify the four model versions. Labels and leader lines identify lead times of 24, 48, 72, 120, and 168 h. The ERA5 P75 threshold is used for all displayed parameters. Curved contours indicate CSI; straight reference lines indicate FB. Skill increases toward the upper-right corner. Complete eight-parameter diagrams are provided in \href{https://github.com/Mixtures-0218/Examine-ML-based-Models-Tornadic-Environment-Prediction-Performance/tree/f52f9474343860cebb6028e6d2878356a9c7fc3e/figures/supplementary}{Supplementary Figs.~S1 and S2 in the repository}.}
  \label{fig:performance}
\end{figure}

Figure~\ref{fig:performance} shows the results for metrics such as POD, CSI, SR, and FB of the ML-based models across the ten tornado event environments in the performance diagram. The performance of the ML-based models varies across different tornado-relevant parameters.

Specifically, for CAPE prediction in Fig.~\ref{fig:performance}(a) at 24- to 72-hour lead times, Pangu-Weather and FuXi Weather show high POD values at the expense of low SR and lie above the 1.25 FB line, indicating a clear overprediction bias at lead times up to 72 h. In contrast, FengWu V1 underperforms in CAPE prediction, showing markedly lower POD but maintaining slightly higher SR values, mostly below FB = 1, indicating a slight underprediction trend. Note that FengWu V2 is not designed to take ERA5 as an input source, and hence some of the results depicted here should be interpreted with caution. This mismatch limits interpretation of the comparison as a ranking of operational performance.

For CIN (not shown; repository Supplementary Fig.~S1b), model performance remains uniformly subdued across all lead times, showing inconsistent fluctuation rather than a systematic decline. This may originate from the fact that high CIN magnitudes are inherently volatile and prone to large amplitude swings when capping inversions are broken, making them particularly challenging for models to forecast.

The K-Index forecasts in Fig.~\ref{fig:performance}(b) show that all models in all lead times fall below the FB = 1 line, indicating a systematic underprediction of high K-Index values despite relatively high SR scores. Moreover, at longer lead times, the forecasts increasingly regress toward their internal statistical mean, effectively smoothing out extreme K-Index events to produce a nearly zero score for CSI at 168-hour lead time. FengWu V2, in this comparison, is outperformed by all other models at 24--120 hours. Some trends may appear to ``back-bend'' due to noisiness in the small event sample.

Pangu-Weather has the highest POD for positive SRH across all evaluated lead times. The two versions of FengWu continue to underpredict the high positive SRH values at 24--120 h, and their performance falls behind other models, as shown in other subgraphs in Fig.~\ref{fig:performance}. FuXi Weather's predictions at 24--48 h lead times maintain POD, but the success ratio drops substantially. This indicates a substantial increase in the false alarm ratio at the 48-hour lead time, whereas the model remains relatively conservative at the 24-hour lead time.

For the cloud base-relevant parameters LCPL and PLFC (Fig.~\ref{fig:performance}(d,e)), FuXi Weather shows a marked tendency toward overprediction relative to the other models. This overprediction produces an elevated POD, widening the performance gap on this score to about 0.55 relative to FengWu V2 for PLFC at 120 h. Meanwhile, the corresponding SR is lower than that of the other models at 24--48 h. This imbalance results in a high false alarm ratio for FuXi Weather. In the comparison of LCPL and PLFC, all models demonstrate greater skill in predicting high LCPL values than predicting high PLFC values. In the LCPL case, the two FengWu models achieve CSI values of up to 0.42 at short lead times, outperforming Pangu-Weather. The failure of Pangu-Weather in short lead times for LCPL is systematic, and this unique feature will be discussed in Section~\ref{sec:pangu_lcl}.

Finally, for the composite thermodynamic-kinematic parameter STP (Fig.~\ref{fig:performance}(f)), both FuXi Weather and Pangu-Weather attain a similar CSI in 24- to 72-hour lead times. However, FuXi Weather shows an overprediction trend, which originates from the high bias in CAPE prediction. In addition, both FengWu models underpredict high STP values at 24--72 h. Both FengWu models appear to revert toward their internal statistical equilibrium once the forecast lead time exceeds approximately 72 hours, as evidenced by the rapid decline in CSI beyond this threshold. Although all models perform ineffectively at the 168-hour lead time, FuXi and Pangu-Weather still show higher POD than the two FengWu models in STP prediction.

\subsection{Fractions Skill Score}

\begin{figure}[!htbp]
  \centering
  \includegraphics[width=0.95\textwidth,height=0.81\textheight,keepaspectratio]{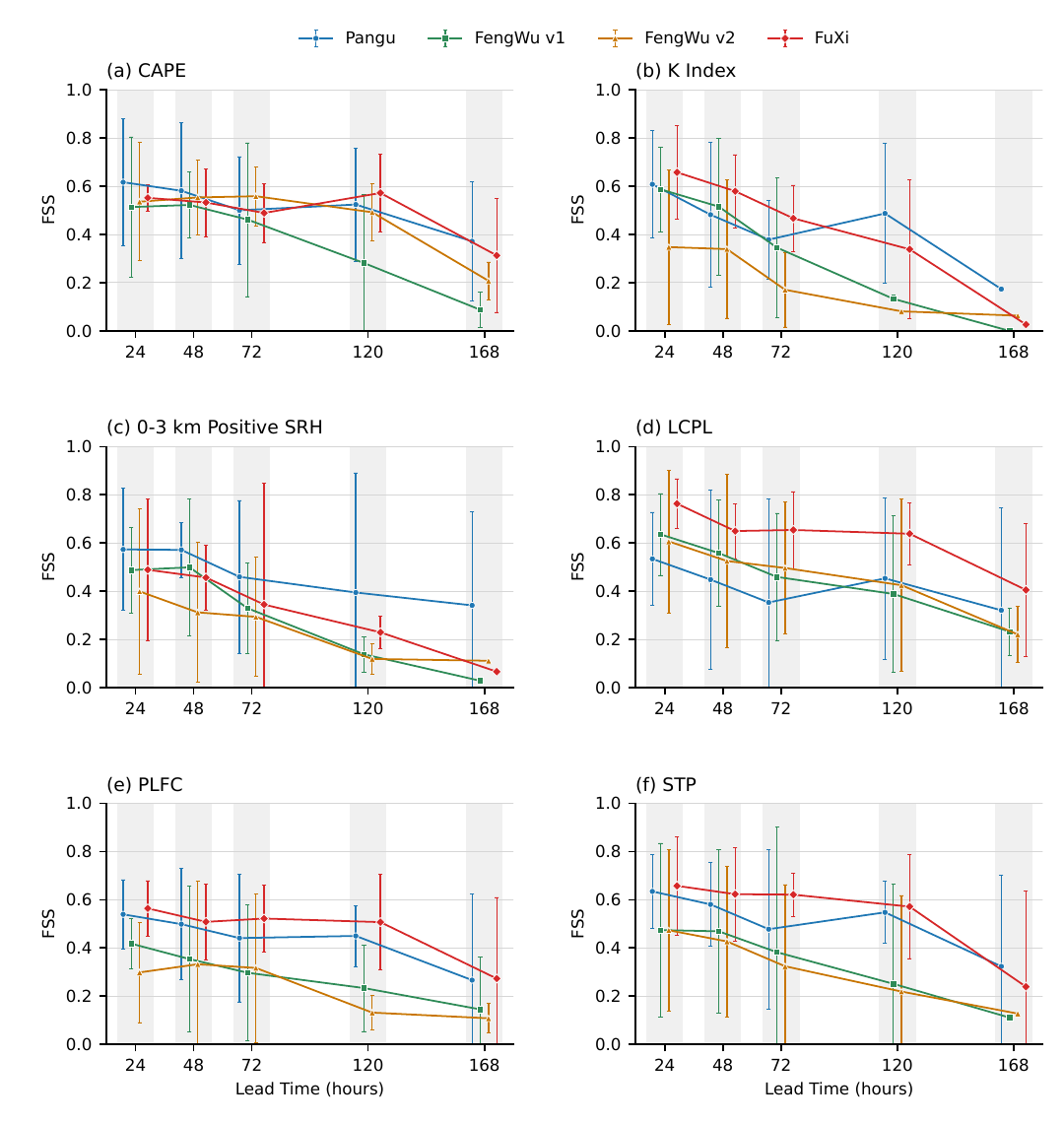}
  \caption{Mean FSS for the same six parameters and four model versions as in Fig.~\ref{fig:performance}, averaged across ten tornado events. The ERA5 P75 threshold is used for all displayed parameters. The neighborhood is $3\times3$ grid points. Colors and marker shapes identify model versions; the horizontal axis shows lead time. Within pale gray bands, model markers are offset horizontally for readability; the bands are display groups, not forecast-validity intervals. Error bars extend by one selected cross-event dispersion scale on either side of the arithmetic mean (Appendix~\ref{app:dispersion}), with portions outside 0--1 clipped. They show event-to-event variability, not uncertainty in the mean. Higher FSS indicates better neighborhood-scale agreement.}
  \label{fig:fss}
\end{figure}

Unlike the point-based metrics in the performance diagrams, FSS is an area-based metric that compares the fractional coverage of threshold-exceeding values within a neighborhood window. Point-wise comparisons would treat even minor spatial displacements as errors: a high-value region offset by one grid cell from the reference produces a miss and a false alarm at each misplaced cell, so a single displacement error is counted twice. Because the high-value regions of tornado-relevant fields are spatially coherent and their precise positions vary slightly with model resolution and smoothing, an area-based metric is better suited here. It credits forecasts whose structure and coverage resemble the reference even when individual features are offset, without inflating the error by double-counting.

In Fig.~\ref{fig:fss}, all models' FSS results show a decreasing trend as lead time increases, indicating that the models progressively smooth out extreme weather events as they revert toward their internal statistical equilibrium. However, the FSS for longer lead times could be slightly greater than that for shorter lead times, which might create an illusion of better model performance at longer lead times~\citep{antonioHowDeriveSkill2025}. This is because FSS can be inflated by forecast bias, and applying the same ERA5 percentile threshold to both fields does not remove forecast frequency bias or the variance deficit caused by over-smoothing. Complete eight-parameter FSS results are provided in \href{https://github.com/Mixtures-0218/Examine-ML-based-Models-Tornadic-Environment-Prediction-Performance/blob/f52f9474343860cebb6028e6d2878356a9c7fc3e/figures/supplementary/S3_fss.pdf}{Supplementary Fig.~S3}.

For CAPE (Fig.~\ref{fig:fss}(a)), short-range FSS values are near 0.6, with Pangu-Weather attaining the highest scores at the 24- and 48-hour lead times. FengWu V2 also maintains a better spatial structure than FengWu V1 at short lead times, consistent with the behavior shown in the performance diagram (Fig.~\ref{fig:performance}(a)).

For the K Index, FuXi Weather outperforms all other models at 24--72 hours, whereas Pangu-Weather scores higher at 120--168 hours. FengWu V1 performs competitively with Pangu-Weather between the 24- and 72-hour lead times, whereas FengWu V2 falls well behind the other models. However, both FengWu models show a rapid decline in FSS with increasing lead time, a pattern that is also evident in the performance diagram (Fig.~\ref{fig:performance}(b)). The FSS generally decreases more rapidly for K Index forecasts than for CAPE, and the spread among models is generally larger. This may be because the K Index is a compound index that combines temperature and moisture information at multiple pressure levels, so the forecast errors of individual layers accumulate as the fields become increasingly smoothed at longer lead times.

For the kinematic parameter (0--3 km positive SRH; Fig.~\ref{fig:fss}(c)), Pangu-Weather achieves an FSS greater than 0.5 at the 24-hour lead time, and this skill level is maintained until the 48-hour lead time. The FSS for this kinematic parameter begins to decline at short lead times, including between 24 and 48 hours for FengWu V2 and FuXi Weather. This early decrease may be associated with smaller-scale rotational features, which exhibit a faster error growth. Overall, Pangu-Weather outperforms the other models for this kinematic parameter and maintains a clear advantage at longer lead times.

FuXi Weather continues to outperform other models in the two cloud-relevant thermodynamic parameters, LCPL and PLFC, in Fig.~\ref{fig:fss}(d) and (e), with FSS of about 0.76 for LCPL and 0.56 for PLFC at 24 hours. The two FengWu model variants show broadly equivalent performance while the broad error bars across all models indicate sensitivity to individual events. Both FengWu versions and FuXi Weather produce reduced FSS for the PLFC forecast relative to the LCPL; Pangu-Weather is an exception at 24--72 hours. In many convective environment fields, parcels in certain areas fail to reach the LFC, which has a substantial effect on model performance. In the PLFC prediction, Pangu-Weather is back to its typical rankings among the models, with forecast skill similar to FuXi Weather but better than the two FengWu models.

In the STP forecast shown in Fig.~\ref{fig:fss}(f), FuXi Weather and Pangu-Weather have a similar ability to capture this composite thermodynamic-kinematic parameter. However, the two FengWu models' FSS quickly decline after a 48-hour lead time. This rapid decline is driven by the drop in skill for 0--3 km positive SRH and CAPE. The gap between FengWu and the top model, FuXi Weather, at the 120-hour lead time reaches approximately 0.32--0.35. Concurrently, Pangu-Weather and FuXi Weather show comparable performance, but FuXi Weather demonstrates a marginal advantage, as its performance barely drops before the 168-hour lead time.

\subsection{Method for Object-based Diagnostic Evaluation}

\begin{figure}[!htbp]
  \centering
  \includegraphics[width=\textwidth]{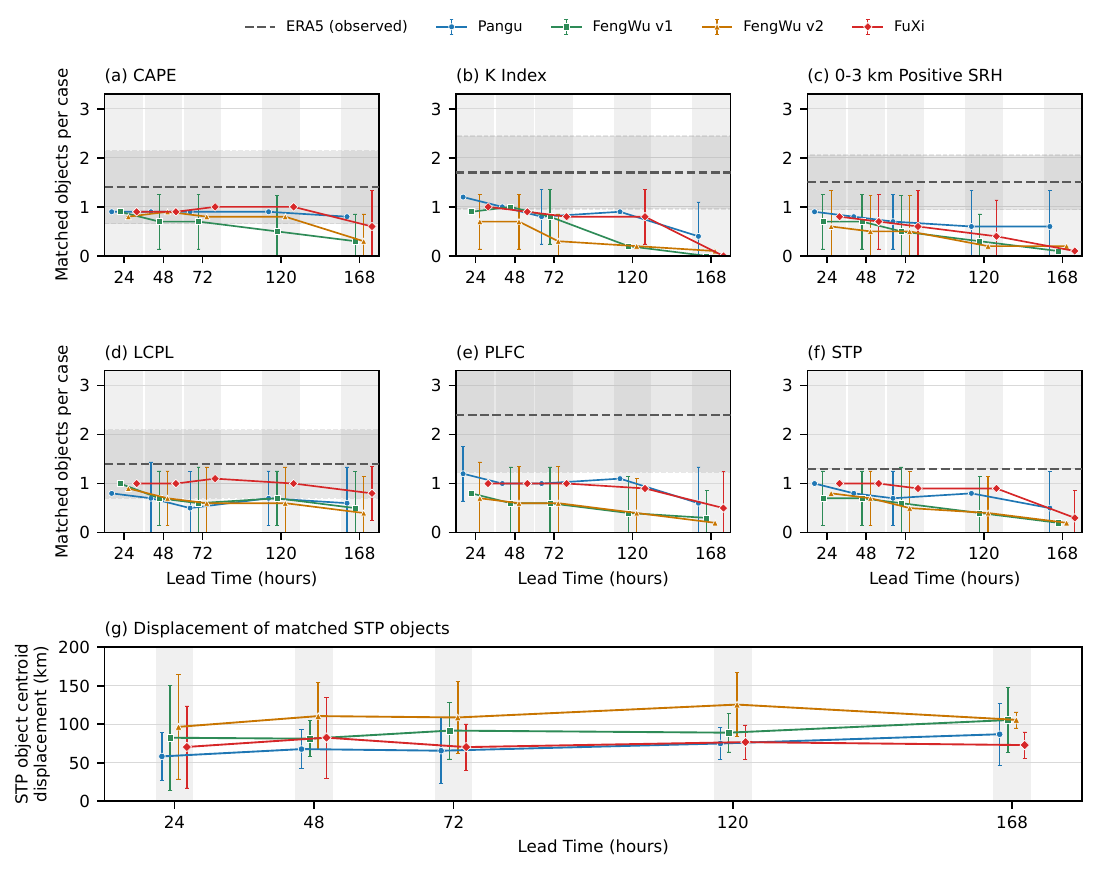}
  \caption{Number of ERA5-matched high-value objects for six tornado-relevant parameters: (a) CAPE, (b) K Index, (c) 0--3 km positive SRH, (d) LCPL, (e) PLFC, and (f) STP. Objects are the connected components of each parameter's binary event field (Appendix~A.3), and a forecast object is counted only if it is paired with an ERA5 object by the one-to-one assignment and the pair reaches a total interest of 0.7; unmatched forecast objects are therefore excluded, and the counts measure how many of the observed high-value structures each model reproduces. The dashed line and shaded band show the ERA5 object count (mean and one selected cross-event dispersion scale, Appendix~\ref{app:dispersion}); error bars are omitted where the scale-selection procedure returns zero, which occurs for concentrated count distributions. (g) Centroid displacement of matched STP objects; because the number of events containing a matched STP object decreases from seven to ten at 24~h to two to five at 168~h, the cross-event dispersion at the longer lead times is based on fewer cases.}
  \label{fig:mode}
\end{figure}

MODE complements FSS by providing spatial verification based on identified objects' localized features. By objectively identifying and characterizing high-value structures in the model fields, MODE can assist operational forecasters in translating ML-based model guidance into practical classification and prediction decisions~\citep{davisMethodObjectBasedDiagnostic2009}. Complete eight-parameter MODE total interest results are provided in \href{https://github.com/Mixtures-0218/Examine-ML-based-Models-Tornadic-Environment-Prediction-Performance/blob/f52f9474343860cebb6028e6d2878356a9c7fc3e/figures/supplementary/S4_mode.pdf}{Supplementary Fig.~S4}. Since FSS and MODE are calculated from the same binary-event field, the MODE total interest pattern by lead time (Supplementary Fig.~S4) is similar to the FSS results, as are the relative rankings of the ML-based models. However, MODE can provide additional information on the actual structure of the identified objects, including the distance deviation, area ratio, and orientation. Figure~\ref{fig:mode} shows the number of matched objects averaged over the ten events per ML-based model for the six main parameters, and how this number varies with lead time. All the models identified fewer objects than the ERA5 baseline. Note that the number of matched objects is inherently sensitive to the thresholds used in the MODE calculation method (the percentile threshold and the 0.7 matching threshold), as described in Appendix~A.3.

Among the parameters shown in Fig.~\ref{fig:mode}, the number of matched objects for CAPE is the closest to the ERA5 baseline. However, the high matching rate of CAPE objects may result from the unique features of CAPE itself and the systematic overprediction in its area: CAPE is a vertically integrated quantity that is inherently more connected than easily fragmented field quantities such as LCPL, positive SRH, and the K Index. As a counterexample, the lower number of matched objects for PLFC is mainly due to parcels not having enough buoyancy to reach the LFC, hence producing separate, multiple, but small objects that are not counted as objects by the MODE algorithm used here. Furthermore, the systematic overprediction in CAPE discussed in Section~\ref{sec:results_perf} could allow the predicted CAPE to reach above the P75 threshold and the 0.7 matching threshold in MODE, further increasing the connectivity by connecting objects and matching with the objects in the reference field.

For the instantaneous field of positive SRH, as shown in Fig.~\ref{fig:mode}(c), the number of matched objects is relatively far from the number of objects in the reference field. However, Pangu-Weather has the highest number of matched objects at most lead times, whereas FuXi Weather usually leads in the thermodynamic parameters. This comparison between FuXi Weather and Pangu-Weather is consistent with~\citet{zhangPhysicsbasedModelsOutperform2026}, who found that FuXi Weather produced lower RMSE for extreme 2m temperature. In contrast, Pangu-Weather performed better for extreme wind events. For STP, Pangu-Weather and FuXi Weather produce comparable results, whereas the two FengWu models systematically underperform relative to both.

MODE also provides the centroid distance deviation of the identified objects from the reference objects, which FSS cannot deliver. For STP, the composite parameter directly relevant to tornado events in Fig.~\ref{fig:mode}(g), Pangu-Weather and FuXi Weather have the smallest average distance deviation. One interesting finding in Fig.~\ref{fig:mode}(g) is that the error bars become unreliable at longer lead times as fewer matched objects are available, and the distance deviation for matched objects increases from 75.4~km to 89.7~km on average over all matched events.

\section{Discussion}
\subsection{Oversmoothing}

\begin{figure}[htbp]
  \centering
  \includegraphics[width=0.95\textwidth,height=0.75\textheight,keepaspectratio]{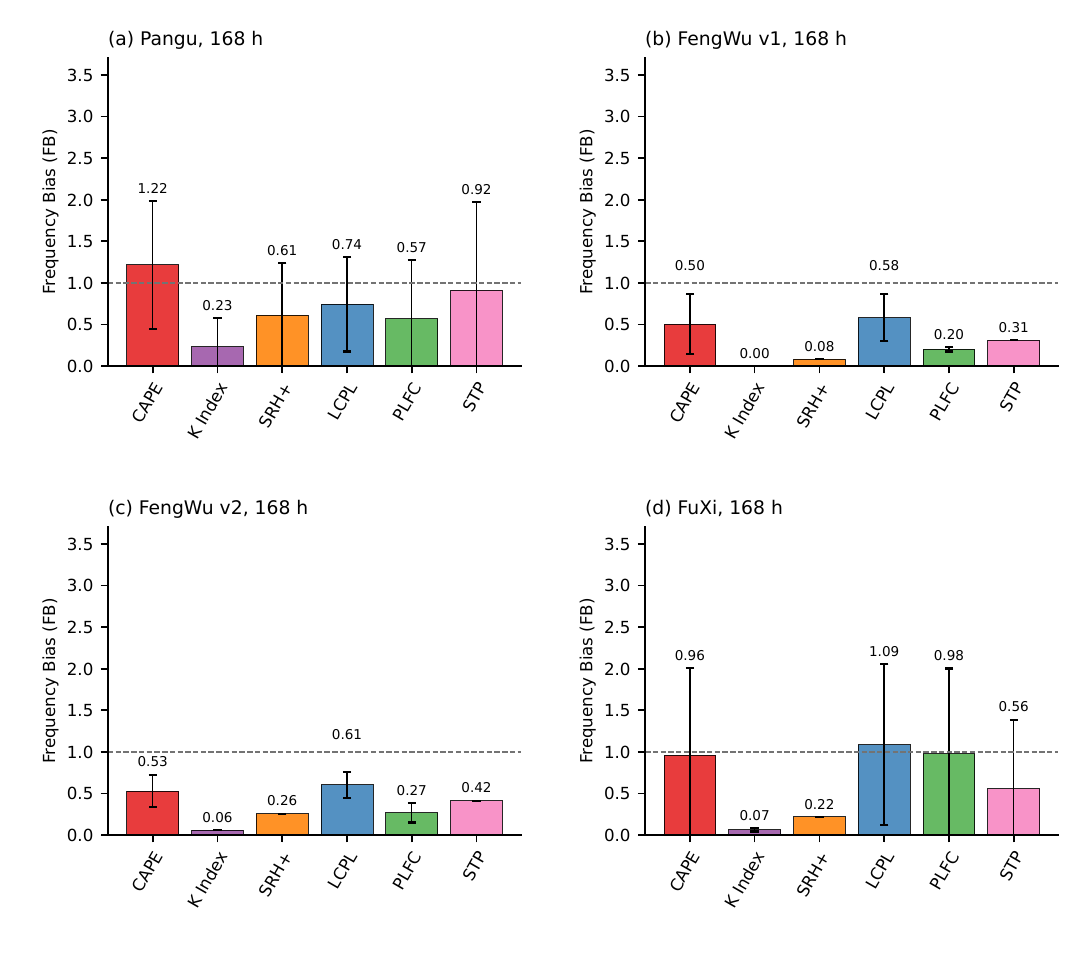}
  \caption{Mean frequency bias of the six main-text parameters for each ML-based model across ten tornado event environments at a lead time of 168 h. Error bars show the mean $\pm$ one cross-case dispersion scale selected using Appendix~\ref{app:dispersion}, with portions below zero clipped. A zero robust scale can occur when many case-level values are identical, even if a few cases differ. The dashed line denotes FB = 1.}
  \label{fig:fb168h}
\end{figure}

The results show that loss of high-end forecast signals depends on both model and parameter. Oversmoothing and mean bias can affect threshold-based event coverage differently, so overprediction alone does not establish oversmoothing. Training with point-wise error losses can favor smoother fields~\citep{bonavitaLimitationsCurrentMachine2024}, but the present comparisons do not isolate the causal contribution of each model's loss function. Figure~\ref{fig:fb168h} summarizes frequency bias at 168 h across the ten events; individual events can depart from these cross-event patterns.

As shown in Fig.~\ref{fig:fb168h}, all models show a tendency to suppress high-magnitude K Index and positive 0--3 km SRH values at longer lead times. For negative 0--3 km SRH, the mean frequency bias is below 1 for all four models when the P25 lower-tail threshold is applied. Both FengWu V1 and FengWu V2 show a pronounced tendency to overpredict CIN while simultaneously diminishing other tornado-favorable signals (\href{https://github.com/Mixtures-0218/Examine-ML-based-Models-Tornadic-Environment-Prediction-Performance/blob/f52f9474343860cebb6028e6d2878356a9c7fc3e/figures/supplementary/S5_frequency_bias.pdf}{Supplementary Fig.~S5}). For negative SRH, the ERA5 P25 threshold equals zero in two of the ten events, making the entire non-positive field an event under the inclusive threshold rule; these cases do not discriminate forecast skill for strong negative SRH. The same limitation applies to the negative-SRH panels in Supplementary Figs.~S2--S4.

The oversmoothing behavior at longer lead times is particularly evident in the two FengWu models, with none of the evaluated parameters remaining close to the FB = 1 benchmark. Moreover, the resulting biases are generally oriented toward less favorable tornado environments, indicating a systematic tendency to suppress ingredients associated with tornado development and return to the climatological mean. Pangu-Weather demonstrates relatively high skill in forecasting the kinematic parameter positive 0--3 km SRH, making it the only model whose MODE total interest for this parameter remains above 0.5 at the 168-hour lead time (Fig.~\ref{fig:mode}). Cross-case dispersion varies across parameters, with some groups having zero robust scales because their values are concentrated at zero (Appendix~\ref{app:dispersion}).

\subsection{CAPE, LCPL, and PLFC overprediction in FuXi Weather}
\label{sec:fuxi_bias}

\begin{figure}[htbp]
  \centering
  \includegraphics[width=\textwidth]{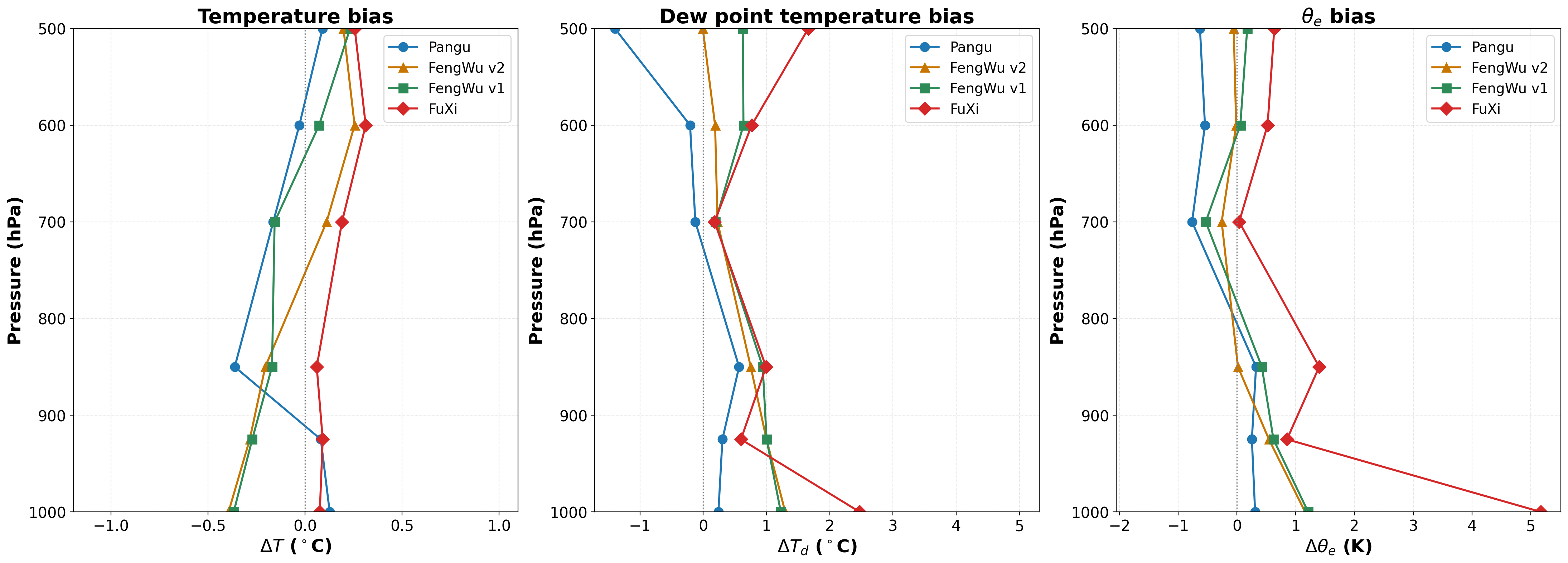}
  \caption{Mean vertical profile deviations from ERA5 (1000--500 hPa) for ML-based model 24 h forecasts, averaged across the ten tornado events.}
  \label{fig:profiles}
\end{figure}

As discussed in Section~\ref{sec:results_perf}, FuXi Weather substantially overpredicts CAPE, LCPL, and PLFC. While it achieves a high probability of detection, its success ratio is relatively low, indicating a high false alarm ratio and, consequently, a significantly larger frequency bias than that of the other ML-based models. These verification metrics suggest that FuXi Weather tends to expand regions of high parameter magnitude into areas where lower values would be expected in the ERA5 reanalysis field. To investigate this unique trend, the mean deviations of vertical profiles between ERA5 and the 24-hour ML-based model forecasts, averaged over the ten tornado event environments over the study region, are shown in Fig.~\ref{fig:profiles}.

As shown in Fig.~\ref{fig:profiles}, the temperature biases are relatively small. However, FuXi Weather shows a pronounced moist bias in the lower troposphere, with a dewpoint temperature bias of +2.39 K at 1000 hPa. Compared with the other models, Pangu-Weather shows only a modest dewpoint temperature bias of +0.22 K at 1000 hPa, whereas the corresponding biases reach +1.17 K in FengWu V1 and +1.24 K in FengWu V2, albeit all with p-values higher than 0.05. When the temperature and dewpoint temperature biases are combined, the resulting equivalent potential temperature bias for FuXi Weather is substantial.

The parameters most strongly affected by this bias---CAPE, LCPL, and PLFC---are highly sensitive to boundary-layer moisture conditions, and the overestimation of dewpoint temperature likely contributes substantially to their forecast biases. In contrast, a similar overprediction trend is not observed for the K Index, which is primarily determined by thermodynamic conditions in the middle troposphere, or for SRH, which is largely independent of atmospheric moisture. These results suggest that the positive biases in the CAPE, LCPL, and PLFC forecasts primarily originate from excessive lower-tropospheric water vapor predicted by FuXi Weather. Although these three parameters are highly affected by the unexpected moist surface environment, the biases are explained by different mechanisms. LCPL is strongly related to the difference between the temperature and dewpoint temperature at 1000 hPa and 925 hPa. The Espy-type approximation relates LCL height to the temperature--dewpoint difference as $z_{\mathrm{LCL}} \approx 125(T-T_d)$, where height is in meters and the temperature difference is in Kelvins~\citep{rompsExactExpressionLifting2017}. Across the ten events, $\Delta(T-T_d)$ at 1000 hPa and $\Delta$LCPL have $r=-0.95$ and $p<0.01$ (not shown).

\begin{figure}[htbp]
  \centering
  \includegraphics[width=\textwidth]{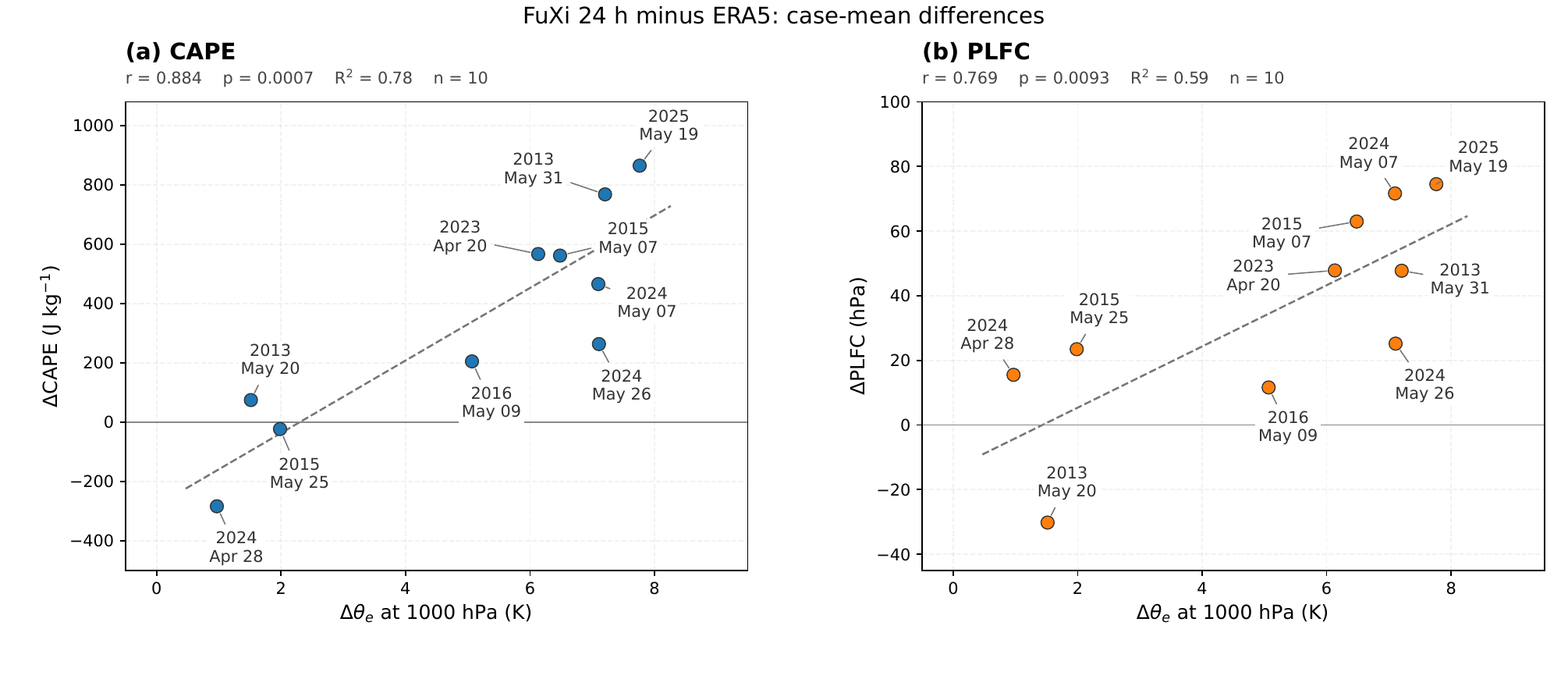}
  \caption{Relationships between the case-mean equivalent potential temperature deviation at 1000 hPa and the case-mean deviations of CAPE and PLFC for the ten tornado event environments. Each point represents the mean FuXi 24-hour forecast minus ERA5 difference over the study region for one environment. Dashed lines show linear fits.}
  \label{fig:cape_lfc_theta_e}
\end{figure}

On the other hand, the deviations in CAPE and PLFC are more related to the surface equivalent potential temperature deviation, with $p<0.01$ (Fig.~\ref{fig:cape_lfc_theta_e}). We used the MetPy~\citep{mayMetPyPythonPackage2025} calculation of dewpoint temperature and equivalent potential temperature.

This surface moisture bias could be the internal bias of the model itself. Although FuXi Weather is the only model that takes in total precipitation (TP), the relationship between TP and CAPE deviation, as well as the surface dewpoint temperature deviation, is not substantial. Furthermore, the large deviation near the tropopause ($\sim$200 hPa; not shown) suggests that ML-based models may show oversmoothing not only in the horizontal dimension but also in the vertical structure of the atmosphere, especially in the higher-level troposphere.

\subsection{Unexpectedly low performance in Pangu-Weather's LCPL prediction}
\label{sec:pangu_lcl}

Pangu-Weather generally ranks among the best-performing models in forecasting most tornado-relevant parameters. However, for LCPL, its forecast skill at lead times from 24 to 72 hours is substantially lower than that of the other ML-based models. It is important to note that all verification metrics presented in Section~\ref{sec:results_perf} are event-based metrics evaluated using the P75 threshold. In contrast, when assessed using continuous metrics such as RMSE, Pangu-Weather achieves the best overall performance among all evaluated models, as shown in Table~\ref{tab:lcl_continuous}.

\begin{table}[htbp]
  \centering
  \caption{Continuous metric comparison among ML-based models in the LCPL 24h lead time predictions over the entire study region across the ten environments. MSE denotes mean squared error.}
  \label{tab:lcl_continuous}
  \small
  \begin{tabular}{@{}lcccc@{}}
    \toprule
    Metrics & Pangu-Weather & FengWu v1 & FengWu v2 & FuXi Weather \\
    \midrule
    Mean Deviation (hPa)           & \textbf{+0.11} & +16.19 & +16.28 & +14.86 \\
    RMSE (hPa)                     & \textbf{37.81} & 50.46  & 56.86  & 50.34  \\
    Spatial Pearson correlation    & \textbf{0.924} & 0.869  & 0.860  & 0.882  \\
    MSE (hPa$^2$)                  & \textbf{2935}  & 3688   & 4230   & 3397   \\
    \bottomrule
  \end{tabular}
\end{table}

The spatial Pearson correlation coefficient is defined in Equation~\ref{eq:pearson} as
\begin{equation}
  \label{eq:pearson}
  r = \frac{\sum_{i=1}^{N}(R_i - \bar{R})(F_i - \bar{F})}
  {\sqrt{\sum_{i=1}^{N}(R_i - \bar{R})^2}\ \sqrt{\sum_{i=1}^{N}(F_i - \bar{F})^2}} ,
\end{equation}
where $R_i$ and $F_i$ denote the reference and forecast values at grid point $i$, respectively, and the overbar denotes the spatial mean. In this case, Pangu-Weather has the highest prediction skill over the entire study region, but has the lowest skill in high-magnitude LCPL prediction.~\citet{bonavitaLimitationsCurrentMachine2024} proposed that ``[ML-based weather forecasting models] tend to perform better on deterministic forecast skill measures.'' The continuous metrics we introduced here could be considered as the deterministic forecast skill measures, and they cannot reflect the actual performance of the model. For example, RMSE cannot distinguish between ``moderate-intensity wide-area oversmoothed fields'' and ``precisely predicted extreme value positions.'' The blurry field is safe in terms of RMSE, but it contains a very limited amount of physical information. Therefore, we break down the LCPL prediction deviation of Pangu-Weather according to three percentile ranges across the entire study region, as shown in Table~\ref{tab:lcl_percentile}.

\begin{table}[htbp]
  \centering
  \caption{LCPL 24--72 h lead time mean prediction deviation (unit: hPa) of Pangu-Weather according to three percentile ranges over the entire study region across the ten environments. P60 and P90 denote the 60th and 90th percentiles, respectively.}
  \label{tab:lcl_percentile}
  \small
  \begin{tabular}{@{}lccc@{}}
    \toprule
    LCPL Percentile range & 24h (hPa) & 48h (hPa) & 72h (hPa) \\
    \midrule
    $<$ P60      & +9.2  & +17.4 & +22.8 \\
    P60 -- P90   & $-9.5$ & $-12.0$ & $-13.5$ \\
    $>$ P90      & $-25.3$ & $-30.3$ & $-36.9$ \\
    \bottomrule
  \end{tabular}
\end{table}

According to Table~\ref{tab:lcl_percentile}, Pangu-Weather shows a pronounced positive bias in regions characterized by low LCPL values, while showing a substantial negative bias in regions with high LCPL values. As the lead time increases, the magnitude of this regression also intensifies. This reflects a fundamental limitation of Pangu-Weather's loss function design, latitude-weighted mean absolute error (MAE) with manually set weights according to surface and pressure level variables~\citep{biAccurateMediumrangeGlobal2023}. Under this architecture, extreme values are systematically drawn toward the model's internal statistical mean, reducing the contrast between high- and low-magnitude regions. While this oversmoothing behavior can improve continuous verification metrics such as RMSE, it also suppresses the representation of extreme events.

\begin{table}[htbp]
  \centering
  \caption{Standard-deviation compression and mean deviation characteristics of 24h lead time LCPL prediction across ML-based models over the study region.}
  \label{tab:variance}
  \small
  \begin{tabular}{@{}lccc@{}}
    \toprule
    Model & Std.-Dev. Compression & Mean Deviation (hPa) & CSI (P75) \\
    \midrule
    Pangu-Weather & $-9.7\%$ & $\sim$0    & 0.351 \\
    FengWu V1     & $-16.4\%$ & +16       & 0.407 \\
    FengWu V2     & $-18.4\%$ & +16       & 0.417 \\
    FuXi Weather  & $-1.0\%$  & +15       & 0.536 \\
    \bottomrule
  \end{tabular}
\end{table}

The pattern shown in Table~\ref{tab:variance} constitutes standard-deviation compression: the LCPL field's spatial standard deviation shrinks to 0.903 of ERA5's (a 9.7\% compression). However, the impact of variance compression on threshold-based metrics like CSI depends critically on whether it is accompanied by a compensating mean bias (Table~\ref{tab:variance}). Pangu-Weather combines a 9.7\% variance compression with a near-zero mean bias ($\Delta$LCPL = +0.1 hPa), the worst-case combination for CSI, since both tails are attenuated with no offsetting shift. FengWu shows even stronger variance compression (16--18\%) but offset by a +16 hPa positive bias, which pushes more grid points above the P75 threshold. FuXi Weather has minimal compression (1.0\%) combined with a +15 hPa positive bias, producing the best CSI.

In contrast, FengWu incorporates an uncertainty loss function through a learnable weighting matrix~\citep{chenOperationalMediumrangeDeterministic2025}, which decreases the weight for meteorological parameters that are more difficult to predict. However, its core remains an L2 loss (mean squared error, which minimizes the squared differences between predictions and targets), so the regression pattern is even more pronounced. Despite this, FengWu overpredicts LCPL, allowing more grid points to exceed the P75 threshold and be counted as successful detections. FengWu also has a diffusion model version called FengWu-ENS~\citep{chenOperationalMediumrangeDeterministic2025} to deal with the disadvantage of uncertainty loss. FuXi Weather employs an L1 loss (mean absolute error, which minimizes the absolute differences between predictions and targets)~\citep{chenFuXiCascadeMachine2023}, but its cascaded forecasting architecture may mitigate some of the limitations inherent to this loss design, and its positive bias further helps overcome the regression trend. Furthermore, FuXi advances forecasts in 6-hour increments, whereas Pangu-Weather uses 24-hour forecast steps. The shorter inference interval in FuXi may help constrain error accumulation between successive forecast cycles. It should also be noted that Pangu-Weather provides models with 1-hour, 3-hour, 6-hour, and 24-hour forecast intervals; however, according to the official implementation guidelines~\citep{xie198808xcPanguWeather2026}, the 24-hour model is the most computationally efficient choice and the default configuration for our study.

As discussed in Section~\ref{sec:fuxi_bias}, LCPL is highly sensitive to near-surface temperature and dewpoint temperature. This characteristic makes LCPL particularly effective at exposing potential limitations in Pangu-Weather's loss function design. The unexpectedly low forecast skill of Pangu-Weather for LCPL likely results from a combination of this parameter's sensitivity and an apparent improvement in CSI among other models arising from their tendency to overpredict favorable conditions. Because Pangu-Weather shows a smaller mean bias in LCPL, the effects of variance compression become more pronounced, leading to a greater reduction in event-based forecast skill despite its better performance in continuous verification metrics. If we manually add a +16 hPa mean bias for Pangu-Weather in the case of Table~\ref{tab:variance}, its CSI value will be raised to 0.472, back to the typical performance of Pangu-Weather in other tornado-relevant parameters.

\subsection{FengWu V2 outperforms FengWu V1 in CAPE prediction}

\begin{figure}[htbp]
  \centering
  \includegraphics[width=0.85\textwidth]{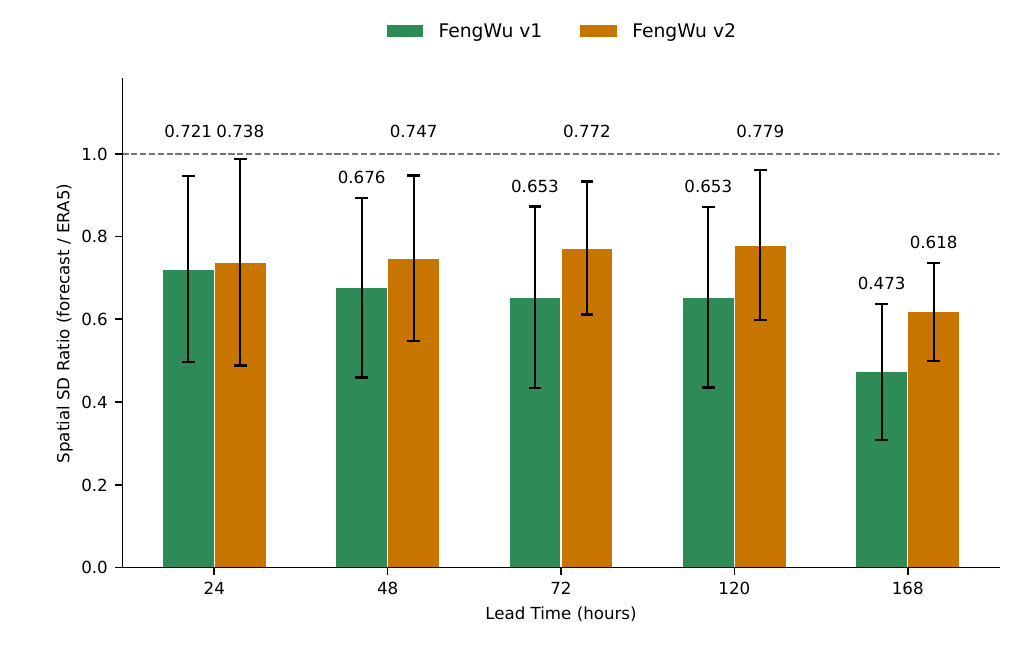}
  \caption{Mean ratio of the forecast to ERA5 spatial standard deviation (``variance preservation'') for FengWu V1 and FengWu V2 over all grid points in the study region. Each case contributes one spatial-SD ratio. Bars show the arithmetic mean across ten cases; error bars extend by one cross-case dispersion scale selected using Appendix~\ref{app:dispersion}. The spatial SD used to define each ratio is distinct from this cross-case dispersion.}
  \label{fig:cape_std}
\end{figure}

It is worth noting that FengWu V2 was developed for operational use rather than for ERA5 and would therefore be expected to perform worse than FengWu V1, which was specifically designed and trained for ERA5 data. This is because ERA5 is a retrospective reanalysis product built on a frozen Integrated Forecasting System (IFS) cycle and is not available in real time, whereas the ECMWF operational analysis is produced in real time by the evolving operational IFS system; to enable real-time deployment, FengWu V2 was further developed through transfer learning from FengWu V1 using ECMWF operational analysis data from 2017--2021~\citep{chenOperationalMediumrangeDeterministic2025}. In this ERA5-based evaluation, FengWu V2 generally underperforms FengWu V1, with CAPE forecasting as the main exception. To further investigate this discrepancy, Fig.~\ref{fig:cape_std} presents the variance preservation rates of CAPE forecasts from both models, averaged across the ten selected environments.

The difference in variance preservation between the two FengWu versions becomes increasingly pronounced as lead time increases. As shown in Fig.~\ref{fig:cape_std} and discussed in Section~\ref{sec:results_perf}, FengWu V2 shows a distinct ``back-bent'' behavior on the performance diagram: its CSI initially increases before decreasing at longer lead times, whereas the CSI of the other models declines in a more nearly linear fashion. This behavior suggests that the short-range forecast skill of FengWu V2 (24--72 h) is partially constrained by the mismatch between its training data and the ERA5 input fields used in this study. Because FengWu V2 was not designed to ingest ERA5 analyses, residual dependence on the initial conditions may negatively affect its early forecast performance.

Nevertheless, FengWu V2 may have learned a more realistic representation of the vertical thermodynamic structure from the operational analysis data on which it was trained. This advantage appears to be particularly beneficial for vertically integrated thermodynamic parameters such as CAPE. In Fig.~\ref{fig:profiles} in Section~\ref{sec:fuxi_bias}, the deviation of the equivalent potential temperature of FengWu V2 is slightly smaller than that of FengWu V1. The subtle advantage could be added up during the integration process when calculating CAPE. As forecast lead time increases beyond approximately 72 hours, the influence of the initial conditions gradually diminishes, and the model's internally learned relationships become increasingly dominant. At these longer lead times, the improved representation of vertical thermodynamic structure in FengWu V2 becomes more apparent, resulting in enhanced variance preservation and better CAPE forecast performance relative to FengWu V1. However, the FengWu research team has not published detailed information regarding the transfer learning procedure used to develop FengWu V2. Consequently, the interpretation presented here should be regarded as a hypothesis rather than a confirmed explanation of the observed performance differences between the two model versions.

\subsection{NWP (GFS) and ML-based model performances in the 2025-05-19 case study}

\begin{figure}[htbp]
  \centering
  \includegraphics[width=\textwidth]{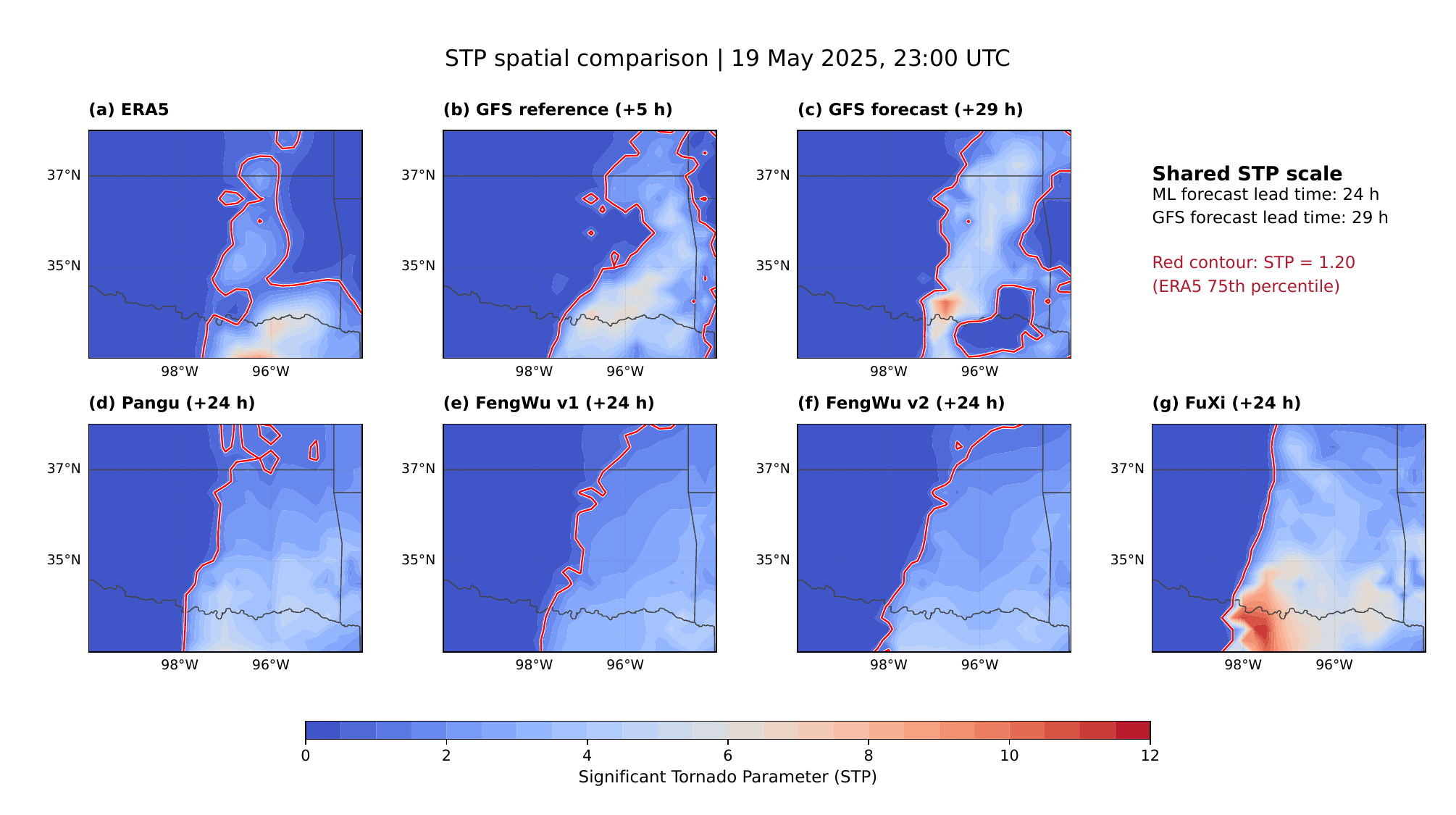}
  \caption{Spatial comparison of STP fields valid at 23:00 UTC on 19 May 2025: ERA5, the GFS reference field, the GFS forecast, and the four ML-based model forecasts. All panels use the same color scale. ML forecast lead times are 24 hours; the actual GFS forecast lead time is 29 hours, and its reference field is a 5-hour forecast. Red contours mark the ERA5 75th-percentile STP threshold in each field.}
  \label{fig:stp24}
\end{figure}

For the comparison in this case study, the GFS is introduced as a benchmark representing the NWP approach. In addition, the ``ground truth'' of GFS will be GFS fct0, and ERA5 will be the ground truth of all ML-based models. Figure~\ref{fig:stp24} presents the STP for the 24-hour lead time predictions from all ML-based models and GFS for the case initiated at 23:00 UTC on 19 May 2025.

All ML-based models in this case produce a markedly smoother prediction field compared to the ERA5 reanalysis. The fine-scale structure near the Red River region ($\sim$34\degr N, 97\degr W) is not resolved by any of the ML-based models, even at a lead time of 24 hours. This unexpectedly low STP value area arises from low CAPE values, which likely indicate the presence of a cold pool. Such fine thermodynamic structures are extremely challenging for statistics-driven ML-based models to capture. In contrast, the GFS forecast in this case successfully depicts a distinct cold pool-like feature in the bottom-right corner. Although the position of this cold pool is slightly shifted east and southward relative to the GFS fct0, it remains evident in the analysis. The core region of STP is more accurately positioned in the GFS forecast, whereas the ML-based models tend to smear the high-end values, thereby smoothing the overall prediction field.

\begin{figure}[htbp]
  \centering
  \includegraphics[width=\textwidth]{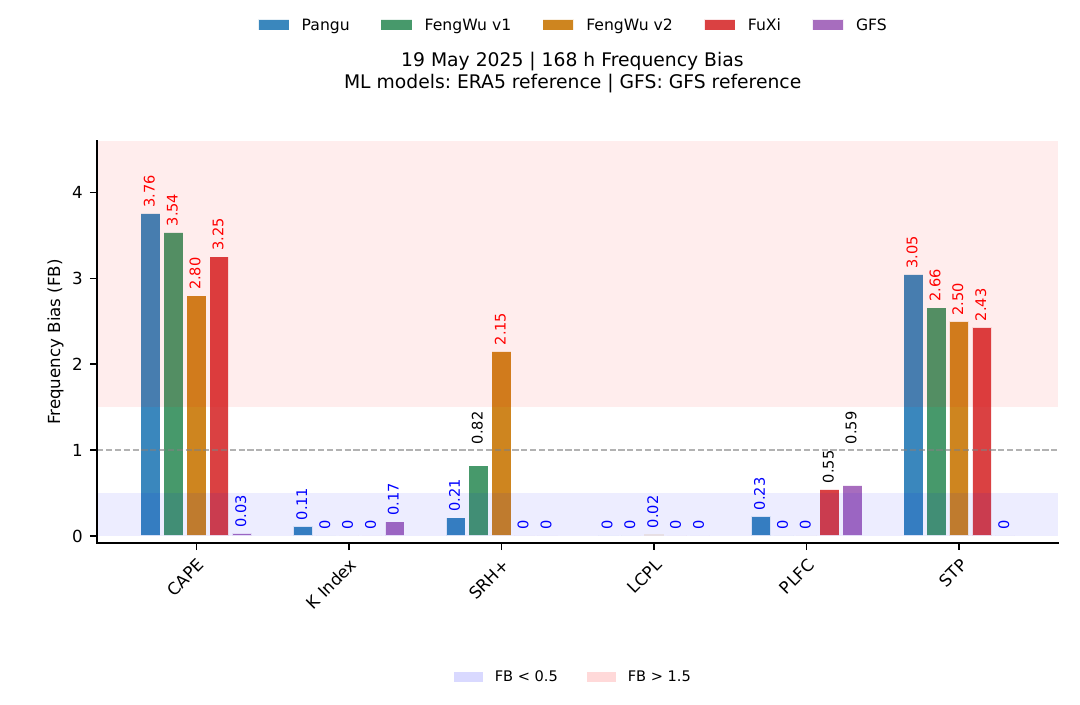}
  \caption{Frequency bias of 168 h forecasts for the six main-text parameters in the 19 May 2025 case. The four ML model versions use ERA5 as the reference; GFS uses its reference forecast. Complete eight-parameter results are shown in \href{https://github.com/Mixtures-0218/Examine-ML-based-Models-Tornadic-Environment-Prediction-Performance/blob/f52f9474343860cebb6028e6d2878356a9c7fc3e/figures/supplementary/S6_case_frequency_bias.pdf}{Supplementary Fig.~S6}.}
  \label{fig:fb168h_case}
\end{figure}

In the long-term (168-hour) forecast, GFS and the ML-based models exhibit contrasting error characteristics in this case, as shown in Fig.~\ref{fig:fb168h_case}: whereas GFS completely loses the convective signal, the ML-based models at least capture a substantial signal of the event at extended lead times, albeit with poor accuracy in both spatial structure and magnitude.

Overall, GFS tends to forecast conditions less favorable for storm development, as it substantially suppresses CAPE and positive SRH to the extent that no STP signal remains in the 168-hour forecast field. Additionally, GFS overpredicts the area of high-magnitude CIN, while its negative-SRH frequency bias is approximately 0.94 (\href{https://github.com/Mixtures-0218/Examine-ML-based-Models-Tornadic-Environment-Prediction-Performance/blob/f52f9474343860cebb6028e6d2878356a9c7fc3e/figures/supplementary/S6_case_frequency_bias.pdf}{Supplementary Fig.~S6}). The performances of GFS and the ML-based models are comparable in predicting K Index, LCPL, and PLFC, as all approaches tend to eliminate the high-end value regions. The ML-based models show a pronounced overprediction trend in CAPE forecasts, in which high-end signals leak into regions that should be low, thereby blurring the overall prediction field and leading to large-scale false alarms. In contrast, GFS displays the opposite tendency, with its forecast field tending to miss the actual event.

\begin{figure}[htbp]
  \centering
  \includegraphics[width=0.9\textwidth,height=0.9\textheight,keepaspectratio]{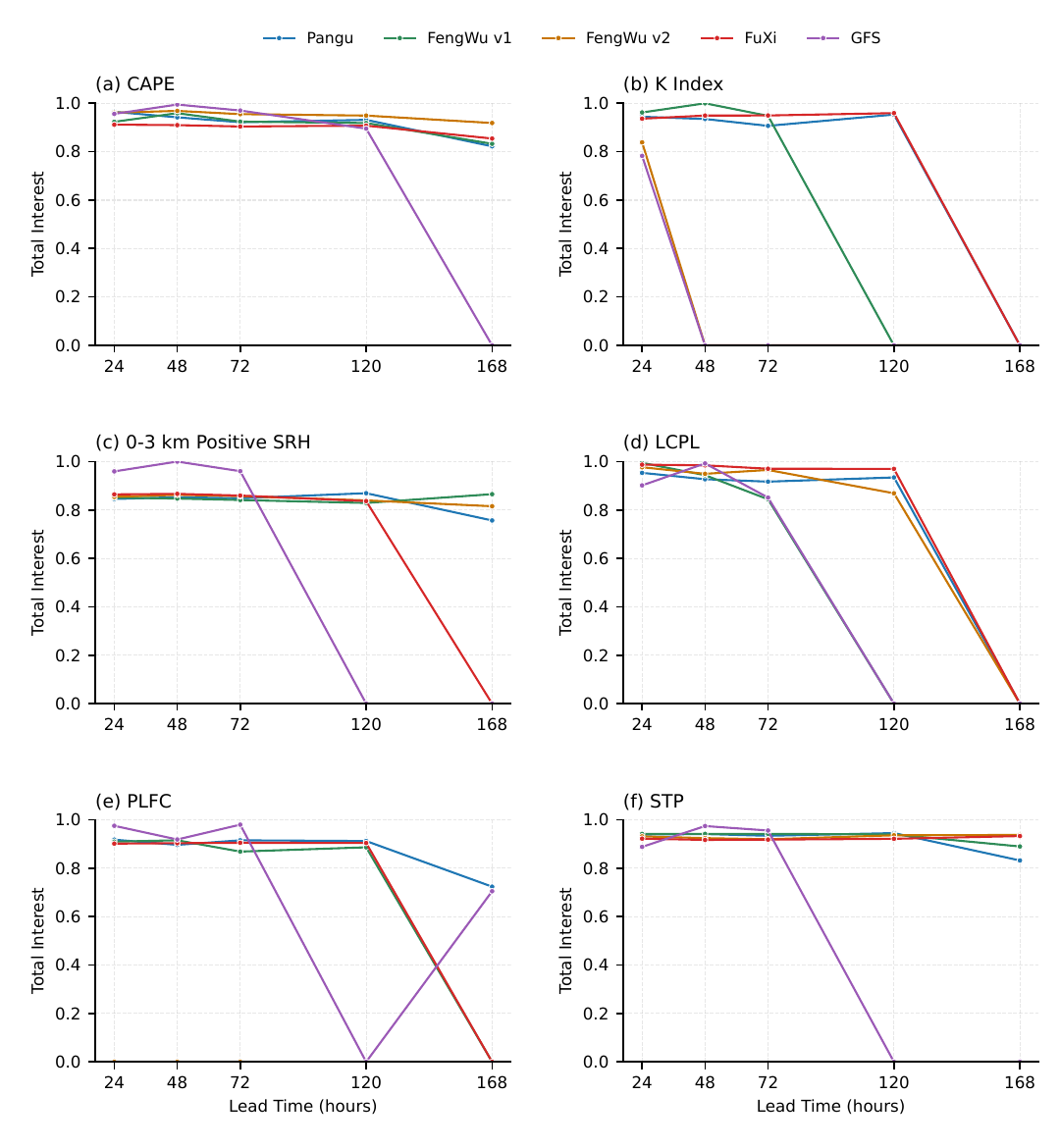}
  \caption{MODE total interest for the six main-text parameters in the 19 May 2025 case. Colors identify the four ML model versions and GFS. The four ML model versions use ERA5 as the reference; GFS uses its reference forecast. Complete eight-parameter results are shown in \href{https://github.com/Mixtures-0218/Examine-ML-based-Models-Tornadic-Environment-Prediction-Performance/blob/f52f9474343860cebb6028e6d2878356a9c7fc3e/figures/supplementary/S7_case_mode.pdf}{Supplementary Fig.~S7}.}
  \label{fig:mode_case}
\end{figure}

The MODE analysis shown in Fig.~\ref{fig:mode_case} further confirms the elimination trend of GFS.

GFS generally maintains a relatively high MODE total interest score for tornado-favorable parameters through 72 hours (0.85--0.98 for CAPE, STP, positive SRH, LCPL, and PLFC). In some instances, the spatial structure of its high-end value areas---for example, in the case of positive SRH---appears more realistic than that of all ML-based models. However, the matched structures collapse at longer lead times: STP, positive SRH, LCPL, and PLFC reach zero at 120 hours, and CAPE at 168 hours, indicating a complete loss of forecast skill. For more challenging parameters such as the K Index, this degradation can occur as early as 48 hours. In contrast, the ML-based models retain MODE scores of 0.82--0.94 for CAPE and STP at 168 hours, although they tend to overpredict certain parameters.

\section{Limitations}
\subsection{Tornado event environment}

Our study includes ten convective environments associated with EF3+ tornadoes. The cross-case error bars summarize variability across these environments rather than uncertainty in the mean (Appendix~\ref{app:dispersion}); distribution diagnostics and dispersion estimates remain sensitive to individual cases in this small sample. The convective environmental configurations are highly diverse, leading to considerable fluctuations in model performance across cases. Nevertheless, given that we selected environments associated with significant tornado production, all models successfully captured the convective features to a substantial degree. Since the study region is confined to central Oklahoma, the generalizability of our findings to other geographical areas may be limited, especially for the coastal regions and tornado environments that are associated with tropical cyclones, as well as high-shear, low-CAPE (HSLC) tornado environments~\citep{sherburnCompositeEnvironmentsSevere2016}. Moreover, the core of the convective environment may not be fully contained within the study domain, potentially resulting in the truncation of critical information, which could adversely affect the performance of the models. Among the ten selected tornado event environments, five occurred outside the training period of all ML-based models, while the remaining five fell within the training period of at least one model. Given the mixture of in-sample and out-of-sample cases, we are unable to distinguish performance differences between the two subsets in our averaged metrics. Nevertheless, this compositional diversity ensures that the evaluation is not overly reliant on either condition. Given that the ERA5 reanalysis provides only hourly data and all ML-based models share an hourly temporal resolution, all convective environments were approximated to the nearest whole hour.

\subsection{NWP (GFS) and ML-based model case study}

The selected convective environments span more than ten years (2013--2025); however, GFS archived high time-resolution prediction data are only available for recent years. Consequently, we are unable to retrieve data for all ten environments to compute the average and perform a comparative evaluation against the ML-based models. In addition, GFS is initialized only four times per day; consequently, the fct0 used as the GFS ground truth, as well as the forecast lead times, may not align precisely with those of the ML-based models. Specifically, we employed the 5-hour forecast initialized at 18:00 UTC on the same day of tornado touchdown as a proxy for the ground truth in the GFS case study. As a result, the actual lead times for GFS correspond to the intended lead time plus 5 hours. For example, the 24-hour lead time is actually a 29-hour lead time in our study. Furthermore, we employed GFS fct0 as the ground truth for GFS specifically, rather than using ERA5, when evaluating the forecast skill of GFS relative to the ML-based models, thereby affording some bias toward higher skill for GFS scores. Nevertheless, this approach is consistent with the inference process of each model: ML-based models, which are initialized with ERA5 data, are compared against ERA5 ground truth, whereas GFS, which relies on its own initialization fields, is evaluated against the output most closely aligned with its data assimilation environment at the corresponding time.

\subsection{Tornado-relevant parameter calculations}

Since all the ML-based models only output basic meteorological parameters, such as temperature, humidity, u/v wind, etc., all the tornado-relevant parameters were calculated from these basic output parameters by using MetPy. To be consistent, the tornado-relevant parameters of ERA5 reanalysis (the ground truth) and the GFS data were also computed from these basic parameters, although they do have direct output of some of the parameters, such as CAPE. In addition,~\citet{wangReanalysesHighResolutionModel2021} suggested that ERA5 may underestimate the ``high tail'' of CAPE, which is attributed to the overly narrow distribution of surface temperature and humidity fields. The vertical wind shear (including 0--6 km bulk wind shear and low-level shear) presented by ERA5 may also be underestimated, particularly for extreme values~\citep{taszarekComparisonConvectiveParameters2021}, and this bias may possibly extend to storm-relative helicity, which could in turn influence our results related to 0--3 km SRH and STP. In this context, neither the initial conditions nor the ground truth provided by ERA5 can fully represent the actual convective environment, nor can GFS fct0, as its output is generated after its own data assimilation process. The shared use of ERA5 for model training, initialization, and verification may inflate apparent forecast skill relative to evaluation against independent observations. In our study, we also find that the ERA5 ground truth and the equivalent GFS fct0 field show substantial differences in both magnitude and the shape of the identified convective objects.

\section{Conclusion}

In this study, we evaluated the performance of three ML-based weather prediction models---Pangu-Weather, FengWu, and FuXi Weather---in reconstructing ten convective environments favorable to strong tornado development. By computing eight tornado-relevant parameters and employing performance diagrams, the Fractions Skill Score (FSS), and the Method for Object-Based Diagnostic Evaluation (MODE), we found that Pangu-Weather and FuXi Weather generally outperformed the two versions of FengWu. Specifically, Pangu-Weather demonstrated a marked advantage in predicting the kinematic parameter positive SRH, whereas FuXi Weather consistently showed the best skill in thermodynamic parameters, albeit with a tendency to overestimate their magnitudes. FengWu V2, a model designed for operational forecasting and built upon FengWu V1, showed the poorest performance in most cases, with the exception of CAPE prediction. All ML-based models, however, suffered from increasing oversmoothing of the prediction field with extended lead times due to their loss function designs, and fine-scale convective structures, such as cold pools, were difficult to capture.

Compared to the ML-based models in our case study, GFS presents a more physically coherent structure for convective environments at the mesoscale, yet rapidly loses its forecast skill when the lead time extends beyond 120 hours. In contrast, the ML-based models are capable of capturing the convective signal, but with poor structural and magnitude fidelity, as they tend to blur the prediction field. The models may achieve higher scores in metrics that average over the entire field, such as RMSE; however, the actual spatial distribution and magnitude of certain tornado-relevant parameters, particularly in the high tail, can deviate substantially from the ground truth. Different models may show their own internal biases; for instance, FuXi Weather systematically overestimates near-surface water vapor content, which in turn leads to substantial biases in CAPE, LCPL, and PLFC forecasts. Overall, the combination of continuous metrics, event-based metrics, and object-based spatial verification provides the most comprehensive characterization to date of the forecast quality of ML-based weather models in tornado-favorable convective environments.

These findings carry important implications for the operational application of ML-based weather models in severe weather forecasting. The systematic oversmoothing of extreme values, combined with model-specific biases in critical thermodynamic fields, suggests that current deterministic ML-based models should be used with caution when assessing tornado event environments at lead times beyond 72 hours. The divergence between continuous metrics (e.g., RMSE) and event-based metrics (e.g., CSI) further emphasizes the need for evaluation frameworks that go beyond traditional global scores when validating ML-based forecasts for high-impact weather. We advocate for the development of open-source evaluation benchmarks specifically designed for convective-scale environments, as well as greater transparency in model training procedures, to facilitate targeted improvements in the next generation of ML-based weather prediction systems.

\appendix
\section{Verification metrics and evaluation domain}
\label{app:metrics}

We evaluate forecast performance over the Oklahoma subdomain. The subdomain considered in this study is bounded by 100\degr W, 94\degr W, 33\degr N, and 38\degr N. At a horizontal resolution of 0.25\degr, this regional domain is represented by a $21\times25$ latitude--longitude grid,
\begin{equation}
  \label{eq:grid}
  N_{\mathrm{lat}} = 21, \qquad N_{\mathrm{lon}} = 25,
\end{equation}
denoting the numbers of latitude and longitude grid points, respectively. In this study, the reference data are obtained from ERA5 reanalysis~\citep{hersbachERA5GlobalReanalysis2020} and are used as the ground-truth dataset for forecast evaluation. For each tornado-relevant parameter $v$, let $\hat{A}^{\,v,t_0}_{i,j}$ denote the ground-truth value at grid point $(i,j)$ valid at time $t_0$, and $A^{\,v,t_0-\tau}_{i,j}$ denote the forecast value of the same parameter at the same grid point, predicted from the earlier time $t_0-\tau$. Here, $v$ is the tornado-relevant parameter, $\tau$ is the lead time, and $t_0$ is the target time to be predicted. The lead times examined in this study are 24, 48, 72, 120, and 168 hours.

Using the ground-truth field $\hat{A}^{\,v,t_0}_{i,j}$, we define a regional threshold $T^v$ as P25 for CIN and negative SRH and P75 for the other parameters. This identifies extreme magnitudes in the lower tail for the two non-positive parameters and in the upper tail for the remaining parameters. For upper-tail parameters, the binary event fields are defined as
\begin{equation}
  \label{eq:binary}
  \hat{B}^{\,v,t_0}_{i,j} =
  \begin{cases}
    1, & \hat{A}^{\,v,t_0}_{i,j} \ge T^v, \\
    0, & \hat{A}^{\,v,t_0}_{i,j} < T^v,
  \end{cases}
  \qquad
  B^{\,v,t_0-\tau}_{i,j} =
  \begin{cases}
    1, & A^{\,v,t_0-\tau}_{i,j} \ge T^v, \\
    0, & A^{\,v,t_0-\tau}_{i,j} < T^v.
  \end{cases}
\end{equation}
For CIN and negative SRH, the inequalities in Eq.~\ref{eq:binary} are reversed: a grid point is an event when its value is at or below the P25 threshold, and otherwise has binary value 0. Thus, binary value 1 identifies the selected tail for each parameter. Here, $\hat{B}^{\,v,t_0}_{i,j} = 1$ means that the ground-truth field identifies an event at grid point $(i,j)$, whereas $B^{\,v,t_0-\tau}_{i,j} = 1$ means that the forecast predicts an event at the same grid point. Under this definition, the numbers of hits, misses, and false alarms can be written in terms of the binary event fields as
\begin{equation}
  \label{eq:hmf}
  H = \sum_{i=1}^{N_{\mathrm{lat}}}\sum_{j=1}^{N_{\mathrm{lon}}} \hat{B}^{\,v,t_0}_{i,j} B^{\,v,t_0-\tau}_{i,j}, \qquad
  M = \sum_{i=1}^{N_{\mathrm{lat}}}\sum_{j=1}^{N_{\mathrm{lon}}} \hat{B}^{\,v,t_0}_{i,j}\left(1 - B^{\,v,t_0-\tau}_{i,j}\right), \qquad
  F = \sum_{i=1}^{N_{\mathrm{lat}}}\sum_{j=1}^{N_{\mathrm{lon}}} \left(1 - \hat{B}^{\,v,t_0}_{i,j}\right) B^{\,v,t_0-\tau}_{i,j}.
\end{equation}

The \textbf{probability of detection (POD)}, the \textbf{Critical Success Index (CSI)}, the \textbf{false alarm ratio (FAR)}, and the \textbf{success ratio (SR)} are defined as
\begin{align}
  \label{eq:pod}
  \mathrm{POD} &= \frac{H}{H + M}, \\
  \label{eq:csi}
  \mathrm{CSI} &= \frac{H}{H + M + F}, \\
  \label{eq:far}
  \mathrm{FAR} &= \frac{F}{H + F}, \\
  \label{eq:sr}
  \mathrm{SR} &= \frac{H}{H + F} = 1 - \mathrm{FAR}.
\end{align}
POD measures the fraction of observed events that are successfully detected by the forecast; CSI provides a composite measure of forecast performance by jointly accounting for hits, misses, and false alarms; FAR measures the fraction of predicted events that did not actually occur; and SR measures the fraction of predicted events that actually occurred. The frequency bias (FB) used in the main text is defined as $\mathrm{FB} = (H + F)/(H + M)$.

\subsection*{A.1 Fractions Skill Score}

While POD, CSI, and SR evaluate threshold-based event detection at individual grid points, the Fractions Skill Score (FSS) assesses whether the forecast reproduces the spatial distribution of tornado-favorable environments over the Oklahoma subdomain. FSS is computed from neighborhood fractions of threshold exceedance. Using the same parameter-specific percentile threshold $T^v$ and tail direction defined above, both the ground-truth field and the forecast field are first converted into binary event fields. In this study, a $3\times3$ neighborhood is used for each central grid point. Then, for each central grid point $(i,j)$, $\hat{O}^{\,v,t_0}_{i,j}$ and $O^{\,v,t_0-\tau}_{i,j}$ denote the local fractions of event grid points within the surrounding $3\times3$ neighborhood in the ground-truth and forecast fields, respectively. Therefore, these two quantities take values between 0 and 1. When the $3\times3$ neighborhood of a boundary grid point extends outside the regional domain, the out-of-domain cells are filled by replicating the value of the nearest boundary grid point. The FSS compares these two neighborhood-fraction fields over the entire regional domain:
\begin{equation}
  \label{eq:fss}
  \mathrm{FSS}(v,\tau) = 1 - \frac{\sum_{i=1}^{N_{\mathrm{lat}}}\sum_{j=1}^{N_{\mathrm{lon}}} \left(\hat{O}^{\,v,t_0}_{i,j} - O^{\,v,t_0-\tau}_{i,j}\right)^2}{\sum_{i=1}^{N_{\mathrm{lat}}}\sum_{j=1}^{N_{\mathrm{lon}}} \left[\left(\hat{O}^{\,v,t_0}_{i,j}\right)^2 + \left(O^{\,v,t_0-\tau}_{i,j}\right)^2\right]} ,
\end{equation}
with values closer to 1 indicating better spatial agreement between the forecast and the ground-truth field.

\subsection*{A.2 Regional root-mean-square error}

For parameters without zero values, such as LCPL and PLFC, we also compute the regional RMSE:
\begin{equation}
  \label{eq:rmse}
  \mathrm{RMSE}(v,\tau) = \sqrt{\frac{\sum_{i=1}^{N_{\mathrm{lat}}}\sum_{j=1}^{N_{\mathrm{lon}}} W(i)\left(\hat{A}^{\,v,t_0}_{i,j} - A^{\,v,t_0-\tau}_{i,j}\right)^2}{N_{\mathrm{lat}} N_{\mathrm{lon}}}} ,
\end{equation}
where $W(i)$ is a latitude-dependent area weight,
\begin{equation}
  \label{eq:w}
  W(i) = N_{\mathrm{lat}} \frac{\cos\phi_i}{\sum_{i'=1}^{N_{\mathrm{lat}}} \cos\phi_{i'}} ,
\end{equation}
and $\phi_i$ is the latitude of the $i$-th grid row, with $i'$ denoting the dummy summation index over all latitude rows. This weight accounts for the fact that grid cells at different latitudes do not represent exactly the same physical area on a spherical Earth.

\subsection*{A.3 Method for Object-Based Diagnostic Evaluation}

MODE is applied directly to the binary event fields defined above. For each tornado-relevant parameter $v$, the thresholded ground-truth field $\hat{B}^{\,v,t_0}$ and forecast field $B^{\,v,t_0-\tau}$ are decomposed into connected spatial objects. Thus, the objects considered by MODE are defined on the basis of the binary event fields rather than on the original continuous-valued parameter fields. Each identified object is then characterized by geometric attributes such as centroid location, boundary separation, area, orientation, and intersection area, and forecast and ground-truth objects are compared through an interest measure that quantifies their similarity.

Following the MET documentation~\citep{20MODETool}, object identification is performed using the convolution-thresholding procedure implemented in MODE. The convolution radius is set to 2.4 grid cells, corresponding to approximately 60 km given the 0.25\degr grid spacing, which is about 25 km in the midlatitudes. In addition, objects with an area smaller than 5 grid cells are treated as noise objects and are discarded before the matching procedure.

Let $\hat{O}^{\,v,t_0}_m$ denote the $m$-th ground-truth object identified from $\hat{B}^{\,v,t_0}$, where each ground-truth object is defined as a maximal connected set of grid cells with binary value 1 in the ground-truth event field $\hat{B}^{\,v,t_0}$. Similarly, let $O^{\,v,t_0-\tau}_n$ denote the $n$-th forecast object identified from $B^{\,v,t_0-\tau}$. For each forecast--ground-truth object pair, the Total Interest is calculated following the MET documentation as
\begin{equation}
  \label{eq:interest}
  I_{\mathrm{total}}^{m,n} = \frac{2.0\, i_{\mathrm{cent}}^{m,n} + 4.0\, i_{\mathrm{bound}}^{m,n} + 1.0\, i_{\mathrm{angle}}^{m,n} + 1.0\, i_{\mathrm{area}}^{m,n} + 2.0\, i_{\mathrm{int\_area}}^{m,n}}{2.0 + 4.0 + 1.0 + 1.0 + 2.0} ,
\end{equation}
where $i_{\mathrm{cent}}^{m,n}$, $i_{\mathrm{bound}}^{m,n}$, $i_{\mathrm{angle}}^{m,n}$, $i_{\mathrm{area}}^{m,n}$, and $i_{\mathrm{int\_area}}^{m,n}$ denote the centroid-distance, boundary-distance, orientation-angle, area-ratio, and intersection-area interest terms, respectively. These interest terms and their corresponding methods are aligned with the MET documentation~\citep{20MODETool}.

To avoid forcing unreasonable matches between dissimilar objects, an assigned pair is regarded as valid only if $I_{\mathrm{total}}^{m,n} \ge 0.7$. If the Total Interest of an assigned pair is smaller than 0.7, that pair is treated as unmatched. Let $\mathcal{P}_{\mathrm{valid}}$ denote the set of valid matched pairs, and let $K = |\mathcal{P}_{\mathrm{valid}}|$ be the number of such pairs. If $K \ge 1$, the mean Total Interest for the case is defined as
\begin{equation}
  \label{eq:mean_interest}
  \overline{I}_{\mathrm{total}} = \frac{1}{K} \sum_{m,n \in \mathcal{P}_{\mathrm{valid}}} I_{\mathrm{total}}^{m,n} .
\end{equation}
Thus, when multiple valid matches exist, the mean Total Interest is taken as the average of their Total Interest values. If no reasonable object match exists, then $\overline{I}_{\mathrm{total}} = 0$.

\section{Cross-case averaging and dispersion}
\label{app:dispersion}

Let $X = \{x_1, x_2, \dots, x_n\}$ contain one metric value from each of the $n=10$ tornado event environments. The center is always the arithmetic mean, $\bar{x}=n^{-1}\sum_{i=1}^{n}x_i$, including when a trimmed or robust dispersion is selected. The following procedure selects a scale for cross-case variability without changing this mean.

We use the Shapiro--Wilk test together with bias-corrected sample skewness and excess kurtosis. Their standardized diagnostics are
\begin{equation}
  \label{eq:zscore}
  Z_{\mathrm{skew}} = \frac{\mathrm{skewness}}{\sqrt{6/n}}, \qquad
  Z_{\mathrm{kurt}} = \frac{\mathrm{excess\ kurtosis}}{\sqrt{24/n}}.
\end{equation}
The denominators are approximate standard errors. With ten cases, these diagnostics guide scale selection rather than establish the population distribution.

\subsection*{B.1 Sample standard deviation}

The sample SD is selected only when all three conditions hold:
\begin{equation}
  \label{eq:normal}
  p \ge 0.05, \qquad |Z_{\mathrm{skew}}| \le 1.96, \qquad |Z_{\mathrm{kurt}}| \le 1.96,
\end{equation}
where $p$ is the Shapiro--Wilk $p$ value. The selected dispersion is then
\begin{equation}
  \label{eq:sd}
  s = \sqrt{\frac{1}{n-1}\sum_{i=1}^{n}(x_i - \bar{x})^2}.
\end{equation}

\subsection*{B.2 Trimmed standard deviation}

For a dataset that fails the joint conditions above, trimmed SD is considered only if
\begin{equation}
  \label{eq:mild}
  1.96 < |Z_{\mathrm{skew}}| \le 3.0, \qquad |Z_{\mathrm{kurt}}| \le 3.0.
\end{equation}
Using linearly interpolated quartiles $Q_1$ and $Q_3$, let $\mathrm{IQR}=Q_3-Q_1$. The retained subset is
\begin{equation}
  \label{eq:outlier}
  X_{\mathrm{trim}}=\{x_i: Q_1-3\,\mathrm{IQR}\le x_i\le Q_3+3\,\mathrm{IQR}\}.
\end{equation}
The sample SD of this subset is selected only if its size $n_{\mathrm{trim}}>0.9n$. Equality at 90\% does not meet this condition. The trimmed SD uses the subset mean and divisor $n_{\mathrm{trim}}-1$; the plotted center remains the mean of all ten cases. Otherwise, the robust scale below is selected.

\subsection*{B.3 Robust dispersion scale}

All datasets that fail the joint conditions in B.1 and do not qualify for trimmed SD in B.2 use the robust scale. This includes datasets rejected only by the Shapiro--Wilk test and those with large standardized skewness or kurtosis. Define
\begin{equation}
  \label{eq:mad}
  \mathrm{MAD}=\mathrm{median}\left(|x_i-\mathrm{median}(X)|\right).
\end{equation}
The selected scale is
\begin{equation}
  \label{eq:robust}
  \sigma_{\mathrm{robust}}=1.4826\,\mathrm{MAD}\quad\mathrm{if}\ \mathrm{MAD}>0,
\end{equation}
and
\begin{equation}
  \label{eq:robust2}
  \sigma_{\mathrm{robust}}=\frac{\mathrm{IQR}}{1.35}\quad\mathrm{if}\ \mathrm{MAD}=0.
\end{equation}
For a normal distribution, $\mathrm{MAD}\approx0.6745\,\sigma$ and $\mathrm{IQR}\approx1.35\,\sigma$; these constants place the estimates on an approximately SD-comparable scale~\citep{rousseeuwAlternativesMedianAbsolute1993, chiTenCircumstancesSolutions2023}. For concentrated or zero-heavy samples, MAD and IQR may both be zero despite a few differing cases. The resulting zero robust scale does not imply identical case-level values.

\subsection*{B.4 Interpretation of error bars}
\label{app:errorbars}

The FSS, MODE, frequency-bias, and CAPE spatial-SD-ratio figures use the same scale-selection procedure above. If the selected scale is $d$, the error bar extends from $\bar{x}-d$ to $\bar{x}+d$, subject to the displayed axis limits. Each arm has length $d$ and the full untruncated span is $2d$. Only the B.1 branch represents one ordinary sample SD on each side. These bars describe cross-case dispersion, not standard errors, confidence intervals for the mean, or intervals with a fixed coverage probability. The spatial SD within each forecast or ERA5 field remains the metric used to define a case-level spatial-SD ratio; only the variability of those ratios across cases follows this appendix.

\clearpage

\section*{Data Availability Statement}
ERA5 reanalysis data used in this study are available from the Copernicus Climate Data Store (\url{https://cds.climate.copernicus.eu})~\citep{hersbachERA5GlobalReanalysis2020}. The GFS 0.25\degr{} historical archive is available from the Research Data Archive at the NSF National Center for Atmospheric Research (doi:10.5065/D65D8PWK) \citep{commerceNCEPGFS0252015}. NASA Shuttle Radar Topography Mission (SRTM) data are available from NASA Earthdata \citep{earthsciencedatasystemsNASAShuttleRadar2025}. Tornado event information was obtained from the NOAA Storm Events Database \citep{StormEventsDatabase}.

\section*{Code Availability}
The code used to compute the tornado-relevant parameters, the verification metrics, and all figures in this study is openly available at \url{https://github.com/Mixtures-0218/Examine-ML-based-Models-Tornadic-Environment-Prediction-Performance} and is archived in Zenodo under DOI \href{https://doi.org/10.5281/zenodo.24481472}{10.5281/zenodo.24481472}~\citep{luoAnalysisCodeTornadic2026}. The MetPy package \citep{mayMetPyPythonPackage2025} and the MET tools are used for the parameter and MODE interest calculations. The complete eight-parameter supplementary figures and their numerical source data are provided in the \href{https://github.com/Mixtures-0218/Examine-ML-based-Models-Tornadic-Environment-Prediction-Performance/tree/f52f9474343860cebb6028e6d2878356a9c7fc3e/figures/supplementary}{repository supplementary collection}.

\section*{Competing Interests}
The authors declare no competing interests.



\begin{thebibliography}{99}
\sloppy

\bibitem[Antonio and Aitchison(2025)]{antonioHowDeriveSkill2025}
Antonio, B., and L. Aitchison (2025), How to derive skill from the Fractions Skill Score, \emph{Mon. Weather Rev.}, \emph{153}(6), 1021--1033, doi:10.1175/MWR-D-24-0120.1.

\bibitem[Bi et al.(2023)]{biAccurateMediumrangeGlobal2023}
Bi, K., L. Xie, H. Zhang, X. Chen, X. Gu, and Q. Tian (2023), Accurate medium-range global weather forecasting with 3D neural networks, \emph{Nature}, \emph{619}, 533--538, doi:10.1038/s41586-023-06185-3.

\bibitem[Bonavita(2024)]{bonavitaLimitationsCurrentMachine2024}
Bonavita, M. (2024), On some limitations of current machine learning weather prediction models, \emph{Geophys. Res. Lett.}, \emph{51}, e2023GL107377, doi:10.1029/2023GL107377.

\bibitem[Bouall\`egue et al.(2024)]{bouallegueRiseDataDrivenWeather2024}
Bouall\`egue, Z. B., et al. (2024), The rise of data-driven weather forecasting: A first statistical assessment of machine learning--based weather forecasts in an operational-like context, \emph{Bull. Am. Meteorol. Soc.}, \emph{105}(6), E864--E883, doi:10.1175/BAMS-D-23-0162.1.

\bibitem[Brooks et al.(2003)]{brooksSpatialDistributionSevere2003}
Brooks, H. E., J. W. Lee, and J. P. Craven (2003), The spatial distribution of severe thunderstorm and tornado environments from global reanalysis data, \emph{Atmos. Res.}, \emph{67--68}, 73--94, doi:10.1016/S0169-8095(03)00045-0.

\bibitem[Burgess et al.(2014)]{burgess20May20132014}
Burgess, D., et al. (2014), 20 May 2013 Moore, Oklahoma, tornado: Damage survey and analysis, \emph{Weather Forecast.}, \emph{29}(5), 1229--1237, doi:10.1175/WAF-D-14-00039.1.

\bibitem[Chen et al.(2025)]{chenOperationalMediumrangeDeterministic2025}
Chen, K., et al. (2025), The operational medium-range deterministic weather forecasting can be extended beyond a 10-day lead time, \emph{Commun. Earth Environ.}, \emph{6}, 518, doi:10.1038/s43247-025-02502-y.

\bibitem[Chen et al.(2023)]{chenFuXiCascadeMachine2023}
Chen, L., X. Zhong, F. Zhang, Y. Cheng, Y. Xu, Y. Qi, and H. Li (2023), FuXi: A cascade machine learning forecasting system for 15-day global weather forecast, \emph{npj Clim. Atmos. Sci.}, \emph{6}, 190, doi:10.1038/s41612-023-00512-1.

\bibitem[Chi et al.(2023)]{chiTenCircumstancesSolutions2023}
Chi, K.-Y., M.-Y. Li, C. Chen, and E. Kang (2023), Ten circumstances and solutions for finding the sample mean and standard deviation for meta-analysis, \emph{Syst. Rev.}, \emph{12}, 62, doi:10.1186/s13643-023-02217-1.

\bibitem[Davis et al.(2009)]{davisMethodObjectBasedDiagnostic2009}
Davis, C. A., B. G. Brown, R. Bullock, and J. Halley-Gotway (2009), The method for object-based diagnostic evaluation (MODE) applied to numerical forecasts from the 2005 NSSL/SPC spring program, \emph{Weather Forecast.}, \emph{24}(5), 1252--1267, doi:10.1175/2009WAF2222241.1.

\bibitem[Feldmann et al.(2024)]{feldmannLightningFastConvectiveOutlooks2024}
Feldmann, M., T. Beucler, M. Gomez, and O. Martius (2024), Lightning-fast convective outlooks: Predicting severe convective environments with global AI-based weather models, \emph{Geophys. Res. Lett.}, \emph{51}(22), e2024GL110960, doi:10.1029/2024GL110960.

\bibitem[Hersbach et al.(2020)]{hersbachERA5GlobalReanalysis2020}
Hersbach, H., et al. (2020), The ERA5 global reanalysis, \emph{Q. J. R. Meteorol. Soc.}, \emph{146}(730), 1999--2049, doi:10.1002/qj.3803.

\bibitem[Hua et al.(2025)]{huaPerformancePanguWeatherDeep2025}
Hua, Z., G. Hakim, and A. Anderson-Frey (2025), Performance of the Pangu-Weather deep learning model in forecasting tornadic environments, \emph{Geophys. Res. Lett.}, \emph{52}, e2024GL109611, doi:10.1029/2024GL109611.

\bibitem[Lang et al.(2024)]{langAIFSECMWFsDatadriven2024}
Lang, S., et al. (2024), AIFS---ECMWF's data-driven forecasting system, preprint, arXiv:2406.01465, doi:10.48550/arXiv.2406.01465.

\bibitem[Lorenz(1969)]{lorenzPredictabilityFlowWhich1969}
Lorenz, E. N. (1969), The predictability of a flow which possesses many scales of motion, \emph{Tellus}, \emph{21}(3), doi:10.3402/tellusa.v21i3.10086.

\bibitem[Luo et al.(2026)]{luoAnalysisCodeTornadic2026}
Luo, Y., R. Yuan, S. Xia, R. Lee, and A. Anderson-Frey (2026), Analysis code for ``Evaluation of Machine-Learning Weather Models in Reproducing Mesoscale Tornado-Favorable Convective Environments'' (v1.0.0) [Computer software], Zenodo, doi:10.5281/zenodo.24481472.

\bibitem[May et al.(2025)]{mayMetPyPythonPackage2025}
May, R. M., S. C. Arms, P. Marsh, E. Bruning, J. R. Leeman, K. Goebbert, J. E. Thielen, Z. Bruick, and M. D. Camron (2025), MetPy: A Python package for meteorological data [Computer software], Unidata, doi:10.5065/D6WW7G29.

\bibitem[Melhauser and Zhang(2012)]{melhauserPracticalIntrinsicPredictability2012}
Melhauser, C., and F. Zhang (2012), Practical and intrinsic predictability of severe and convective weather at the mesoscales, \emph{J. Atmos. Sci.}, \emph{69}(11), 3350--3371, doi:10.1175/JAS-D-11-0315.1.

\bibitem[NASA Earth Science Data Systems(2025)]{earthsciencedatasystemsNASAShuttleRadar2025}
NASA Earth Science Data Systems (2025), NASA Shuttle Radar Topography Mission global 1 arc second V003 [dataset], NASA Earthdata, https://www.earthdata.nasa.gov/data/catalog/lpcloud-srtmgl1-003.

\bibitem[National Center for Atmospheric Research(n.d.)]{20MODETool}
National Center for Atmospheric Research (n.d.), Model evaluation tools (MET) user's guide: MODE tool, Developmental Testbed Center, accessed 3 May 2026, https://metplus.readthedocs.io/projects/met/en/latest/Users\_Guide/mode.html.

\bibitem[National Centers for Environmental Information(2013)]{StormEventsDatabasea}
National Centers for Environmental Information (2013), Storm events database: Event details, NOAA, accessed 26 July 2026, https://www.ncei.noaa.gov/stormevents/eventdetails.jsp?id=451572.

\bibitem[National Centers for Environmental Information(n.d.)]{StormEventsDatabase}
National Centers for Environmental Information (n.d.), Storm events database, NOAA, accessed 26 July 2026, https://www.ncei.noaa.gov/stormevents/.

\bibitem[National Centers for Environmental Prediction(2015)]{commerceNCEPGFS0252015}
National Centers for Environmental Prediction (2015), NCEP GFS 0.25 degree global forecast grids historical archive [dataset], NSF National Center for Atmospheric Research, doi:10.5065/D65D8PWK.

\bibitem[Pathak et al.(2022)]{pathakFourCastNetGlobalDatadriven2022}
Pathak, J., et al. (2022), FourCastNet: A global data-driven high-resolution weather model using adaptive Fourier neural operators, preprint, arXiv:2202.11214, doi:10.48550/arXiv.2202.11214.

\bibitem[Roberts and Lean(2008)]{robertsScaleSelectiveVerificationRainfall2008}
Roberts, N. M., and H. W. Lean (2008), Scale-selective verification of rainfall accumulations from high-resolution forecasts of convective events, \emph{Mon. Weather Rev.}, \emph{136}, 78--97, doi:10.1175/2007MWR2123.1.

\bibitem[Roebber(2009)]{roebberVisualizingMultipleMeasures2009}
Roebber, P. J. (2009), Visualizing multiple measures of forecast quality, \emph{Weather Forecast.}, \emph{24}(2), 601--608, doi:10.1175/2008WAF2222159.1.

\bibitem[Rousseeuw and Croux(1993)]{rousseeuwAlternativesMedianAbsolute1993}
Rousseeuw, P. J., and C. Croux (1993), Alternatives to the median absolute deviation, \emph{J. Am. Stat. Assoc.}, \emph{88}(424), 1273--1283, doi:10.2307/2291267.

\bibitem[Romps(2017)]{rompsExactExpressionLifting2017}
Romps, D. M. (2017), Exact expression for the lifting condensation level, \emph{J. Atmos. Sci.}, \emph{74}(12), 3891--3900, doi:10.1175/JAS-D-17-0102.1.

\bibitem[Sherburn et al.(2016)]{sherburnCompositeEnvironmentsSevere2016}
Sherburn, K. D., M. D. Parker, J. R. King, and G. M. Lackmann (2016), Composite environments of severe and nonsevere high-shear, low-CAPE convective events, \emph{Weather Forecast.}, \emph{31}(6), 1899--1927, doi:10.1175/WAF-D-16-0086.1.

\bibitem[Stensrud et al.(2013)]{stensrudProgressChallengesWarnonForecast2013}
Stensrud, D. J., et al. (2013), Progress and challenges with Warn-on-Forecast, \emph{Atmos. Res.}, \emph{123}, 2--16, doi:10.1016/j.atmosres.2012.04.004.

\bibitem[Taszarek et al.(2021)]{taszarekComparisonConvectiveParameters2021}
Taszarek, M., N. Pilguj, J. T. Allen, V. Gensini, H. E. Brooks, and P. Szuster (2021), Comparison of convective parameters derived from ERA5 and MERRA-2 with rawinsonde data over Europe and North America, \emph{J. Clim.}, \emph{34}(8), 3211--3237, doi:10.1175/JCLI-D-20-0484.1.

\bibitem[Thompson et al.(2003)]{thompsonCloseProximitySoundings2003}
Thompson, R. L., R. Edwards, J. A. Hart, K. L. Elmore, and P. Markowski (2003), Close proximity soundings within supercell environments obtained from the Rapid Update Cycle, \emph{Weather Forecast.}, \emph{18}(6), 1243--1261, doi:10.1175/1520-0434(2003)018<1243:CPSWSE>2.0.CO;2.

\bibitem[Thompson et al.(2004)]{thompsonUpdateSupercellComposite2004}
Thompson, R. L., R. Edwards, and C. M. Mead (2004), An update to the supercell composite and significant tornado parameters, paper presented at 22nd Conf. on Severe Local Storms, Amer. Meteorol. Soc., Hyannis, Mass.

\bibitem[Wang et al.(2021)]{wangReanalysesHighResolutionModel2021}
Wang, Z., J. A. Franke, Z. Luo, and E. J. Moyer (2021), Reanalyses and a high-resolution model fail to capture the ``high tail'' of CAPE distributions, \emph{J. Clim.}, \emph{34}(21), 8699--8715, doi:10.1175/JCLI-D-20-0278.1.

\bibitem[Xie(2026)]{xie198808xcPanguWeather2026}
Xie, L. (2026), 198808xc/Pangu-Weather [Computer software], GitHub, https://github.com/198808xc/Pangu-Weather.

\bibitem[Zhang et al.(2026)]{zhangPhysicsbasedModelsOutperform2026}
Zhang, Z., E. Fischer, J. Zscheischler, and S. Engelke (2026), Physics-based models outperform AI weather forecasts of record-breaking extremes, \emph{Sci. Adv.}, \emph{12}(18), eaec1433, doi:10.1126/sciadv.aec1433.

\end{thebibliography}
\end{document}